# Observation of new microsecond isomers among fission products of 345 MeV/nucleon $^{238}$U

D. Kameda,[1,*] T. Kubo,[1] T. Ohnishi,[1] K. Kusaka,[1] A. Yoshida,[1] K. Yoshida,[1] M. Ohtake,[1] N. Fukuda,[1] H. Takeda,[1] K. Tanaka,[1] N. Inabe,[1] Y. Yanagisawa,[1] Y. Gono,[1] H. Watanabe,[1] H. Otsu,[1] H. Baba,[1] T. Ichihara,[1] Y. Yamaguchi,[1] M. Takechi,[1] S. Nishimura,[1] H. Ueno,[1] A. Yoshimi,[1] H. Sakurai,[1] T. Motobayashi,[1] T. Nakao,[2] Y. Mizoi,[3] M. Matsushita,[4] K. Ieki,[4] N. Kobayashi,[5] K. Tanaka,[5] Y. Kawada,[5] N. Tanaka,[5] S. Deguchi,[5] Y. Satou,[5] Y. Kondo,[5] T. Nakamura,[5] K. Yoshinaga,[6] C. Ishii,[6] H. Yoshii,[6] Y. Miyashita,[6] N. Uematsu,[6] Y. Shiraki,[6] T. Sumikama,[6] J. Chiba,[6] E. Ideguchi,[7] A. Saito,[7] T. Yamaguchi,[8] I. Hachiuma,[8] T. Suzuki,[8] T. Moriguchi,[9] A. Ozawa,[9] T. Ohtsubo,[10] M. A. Famiano,[11] H. Geissel,[12] A. S. Nettleton,[13] O. B. Tarasov,[13] D. Bazin,[13] B. M. Sherrill,[13] S. L. Manikonda,[14] and J. A. Nolen[14]

[1] *RIKEN Nishina Center, RIKEN, 2-1 Hirosawa, Wako, Saitama 351-0198, Japan*

[2] *Department of Physics, University of Tokyo, 7-3-1 Hongo, Bunkyo-ku, Tokyo 113-0033, Japan*

[3] *Department of Engineering Science, Osaka Electro-Communication University, 18-8 Hatsucho, Neyagawa, Osaka 572-8530, Japan*

[4] *Department of Physics, Rikkyo University, 3-34-1 Nishi-Ikebukuro, Toshima-ku, Tokyo 171-8501, Japan*

[5] *Department of Physics, Tokyo Institute of Technology, 2-12-1 Ookayama, Meguro-ku, Tokyo 152-8551, Japan*

[6] *Faculty of Science and Technology, Tokyo University of Science, 2461 Yamazaki, Noda, Chiba 278-8510, Japan*

[7] *Center for Nuclear Study, University of Tokyo, 2-1 Hirosawa, Wako, Saitama 351-0198, Japan*

[8] *Department of Physics, Saitama University, 255 Shimo-Okubo, Sakura-ku, Saitama City, Saitama 338-8570, Japan*

[9] *Institute of Physics, University of Tsukuba, 1-1-1 Ten'noudai, Tsukuba, Ibaraki 305-8571, Japan*

[10] *Institute of Physics, Niigata University, 8050 Ikarashi 2-no-cho, Nishi-ku, Niigata 950-2181, Japan*

[11] *Department of Physics, Western Michigan University (WMU), 1903 W. Michigan Avenue, Kalamazoo, Michigan 49008-5252, USA*

[12] *Gesellschaft fuer Schwerionenforshung (GSI) mbH, 1 Planckstr, Darmstadt 64291, Germany*
[13] *National Superconducting Cyclotron Laboratory (NSCL), Michigan State University (MSU), 640 South Shaw Lane, East Lansing, Michigan 48824-1321, USA*
[14] *Argonne National Laboratory (ANL), 9700 S. Cass Avenue, Argonne, Illinois 60439, USA*

A search for isomeric $\gamma$-decays among fission fragments from 345 MeV/nucleon $^{238}$U has been performed at the RIKEN Nishina Center RI Beam Factory. Fission fragments were selected and identified using the superconducting in-flight separator BigRIPS and were implanted in an aluminum stopper. Delayed $\gamma$-rays were detected using three clover-type high-purity germanium detectors located at the focal plane within a time window of 20 $\mu$s following the implantation. We identified a total of 54 microsecond isomers with half-lives of ~ 0.1 - 10 $\mu$s, including discovery of 18 new isomers in very neutron-rich nuclei: $^{59}$Ti$^{m}$, $^{90}$As$^{m}$, $^{92}$Se$^{m}$, $^{93}$Se$^{m}$, $^{94}$Br$^{m}$, $^{95}$Br$^{m}$, $^{96}$Br$^{m}$, $^{97}$Rb$^{m}$, $^{108}$Nb$^{m}$, $^{109}$Mo$^{m}$, $^{117}$Ru$^{m}$, $^{119}$Ru$^{m}$, $^{120}$Rh$^{m}$, $^{122}$Rh$^{m}$, $^{121}$Pd$^{m}$, $^{124}$Pd$^{m}$, $^{124}$Ag$^{m}$ and $^{126}$Ag$^{m}$, and obtained a wealth of spectroscopic information such as half-lives, $\gamma$-ray energies, $\gamma$-ray relative intensities and $\gamma\gamma$ coincidences over a wide range of neutron-rich exotic nuclei. Proposed level schemes are presented for $^{59}$Ti$^{m}$, $^{82}$Ga$^{m}$, $^{92}$Br$^{m}$, $^{94}$Br$^{m}$, $^{95}$Br$^{m}$, $^{97}$Rb$^{m}$, $^{98}$Rb$^{m}$, $^{108}$Nb$^{m}$, $^{108}$Zr$^{m}$, $^{109}$Mo$^{m}$, $^{117}$Ru$^{m}$, $^{119}$Ru$^{m}$, $^{120}$Rh$^{m}$, $^{122}$Rh$^{m}$, $^{121}$Pd$^{m}$, $^{124}$Ag$^{m}$ and $^{125}$Ag$^{m}$, based on the obtained spectroscopic information and the systematics in neighboring nuclei. Nature of the nuclear isomerism is discussed in relation to evolution of nuclear structure.

---

* E-mail: kameda@ribf.riken.jp

## I. INTRODUCTION

It is known that the study of microsecond isomers is a useful means to investigate the evolution of shell structure and nuclear shape far from stability. One example where this has been applied is in the region of neutron-rich exotic nuclei with atomic numbers $Z$ ~ 28 to 50. Nuclear isomerism is sensitive to microscopic changes of nuclear structure. For instance, in the vicinity of doubly-magic nuclei $^{78}$Ni and $^{132}$Sn, the so-called seniority isomers are generated by the stretched configuration of high-$j$ orbitals such as $g_{9/2}$ and $h_{11/2}$ [1]. These isomers, such as $^{76}$Ni$^m$ [2], $^{78}$Zn$^m$ [3] and $^{130}$Cd$^m$ [4], provide a means to investigate the persistence of shell gaps far from stability. On the other hand, the so-called shape isomers [5], which are generated due to shape coexistence [6], have been observed in the intermediate region between major shells. The well-known shape isomers are those appearing in the neutron-rich exotic nuclei with atomic numbers $Z$ ~ 38 to 42 and neutron numbers $N$ ~ 58 to 61 [6]. For instance, in the case of $^{98}$Sr$^m$ [7] and $^{100}$Zr$^m$ [8], the spherical second $0^+$ states appear at a low excitation energy above the prolate-deformed ground state, and exhibit isomeric transitions. Furthermore other isomers, such as the so-called $K$-isomers [9, 10] and the isomeric transitions between band heads of low-lying deformed states [11], have been also observed in the intermediate region between $N$ = 50 and 82.

Systematic investigations of isomers in neutron-rich exotic nuclei between $^{78}$Ni and $^{132}$Sn are of interest, because not only shape coexistence but also shape transition phenomena have been predicted by several nuclear models [12-14]. Isomer spectroscopy of such exotic nuclei should allow one to test these nuclear models as well as to investigate the evolution of nuclear shape and shape coexistence. Moreover nuclei in this region are also important for the astrophysical r-process [15, 16], since they are predicted to be located on or close to the r-process path. Despite the interest, isomers in these neutron-rich exotic nuclei have not yet been studied in detail due to the difficulties in producing such very rare isotopes far from stability. The present work is the first attempt to investigate them comprehensively.

The production of radioactive isotopes (RI) using a projectile fragment separator [17] provides us with a unique opportunity to observe short-lived isomers with half-lives of a few microseconds, thanks to in-flight separation. The fast isotopic separation and identification of reaction products, which take place in several hundred nanoseconds, allow event-by-event detection of isomeric $\gamma$-rays at the focal plane of the separator with small decay losses in flight. The $\gamma$ decays are observed under low-background conditions after ion implantation. These features allow the search for and highly-selective spectroscopy of new isomers over a wide range of exotic nuclei far from stability. This was first demonstrated by R. Grzywacz et al. in an experiment at GANIL in 1995 [18], and a number of the succeeding experiments [3, 4, 19-24] have provided a wealth of spectroscopic information on the nuclear structure of exotic nuclei. In these experiments not only projectile fragmentation of heavy-ion beams but also in-flight fission of a uranium beam have been used as production reactions to populate isomers. In-flight fission is known to be an excellent mechanism for producing neutron-rich exotic nuclei, having large production cross sections over a wide range of atomic numbers, as demonstrated by experiments at GSI in which more than a hundred new isotopes were identified [25, 26]. The features of in-flight fission also provide us with an opportunity to expand isomer spectroscopy towards the neutron drip-line. In fact several new isomers in the region of mass number $A$ ~ 100 have been recently discovered at NSCL by C. M. Folden et al. using in-flight fission of an 80 MeV/nucleon $^{238}$U beam [27]. However, in spite of the great potential of in-flight fission, its application to isomer spectroscopy has been limited to relatively narrow regions in the nuclear chart due to limited performance of fragment separator and primary uranium beam. To take full advantage of the production mechanism of in-flight fission, its reaction kinematics requires a fragment separator with large ion-optical acceptance, so that efficient collection of fission fragments can be achieved. Fission fragments are produced with much larger spreads in angle and momentum compared to projectile fragmentation. Increased intensity and energy of the primary uranium beam relative to past experiments provides significant benefits as well. These conditions are met in the new-generation in-flight RI beam facilities that have been or are being developed worldwide, allowing dramatic advances in isomer spectroscopy.

One of the new generation of devices, the superconducting in-flight separator BigRIPS [28, 29] was recently developed at RIKEN RI Beam Factory (RIBF) [30]. It is a two-stage separator with large acceptance so that the features of in-flight fission can be exploited. The second stage of the BigRIPS separator is designed to have high momentum resolution which allows excellent particle identification for RI beams [31]. Such capabilities of the BigRIPS separator have been well demonstrated in an experiment to search for new isotopes using in-flight fission of a $^{238}$U beam at 345 MeV/nucleon, in which we discovered 45 new neutron-rich isotopes with atomic numbers ranging from 25 to 56 [32]. The experiment also allowed us to expand isomer spectroscopy towards the neutron drip-line. While searching for new isotopes, we simultaneously observed delayed $\gamma$-rays at the focal plane in order to search for new isomers as well as to perform isomer tagging [32]. As anticipated, a number of isomeric decays were observed including those from new isomers in very neutron-rich exotic nuclei. We obtained new spectroscopic information on not only the new isomers but also on some known, previously reported, isomers. In this paper we report on the experimental results from our comprehensive isomer search among the fission fragments, which was performed using the BigRIPS separator at RIBF at RIKEN Nishina Center.

## II. EXPERIMENT

The experiment was carried out using a $^{238}U^{86+}$ beam accelerated to 345 MeV/nucleon by the cascade operation of the RIBF accelerator complex. The beam intensity was approximately 0.2 particle nA (pnA) on target. The present data on new isomers were recorded during the same runs as the recent search for new isotopes [32]. The fission fragments produced by the in-flight fission of the uranium beam were analyzed and identified in the BigRIPS separator, transported to the focal plane of the ZeroDegree spectrometer [32] and implanted in an aluminum stopper. Delayed $\gamma$-rays were detected using three clover-type germanium detectors within a time window of 20 $\mu$s following the implantation. See Ref. [31] for the layout and configuration of the BigRIPS separator and ZeroDegree spectrometer. The details of the separator settings are summarized in Table I of Ref. [32]. As described in the reference we employed three different separator settings, each of which targeted new isotopes and new isomers in the regions with atomic numbers around 30, 40, and 50, respectively. In the following, these are referred to as the G1, G2 and G3 settings, respectively.

The details about the production, separation and particle identification of fission fragments are given in Refs. [31] and [32]. Here we describe the experimental procedure for the measurement of isomer decays. The first stage of the BigRIPS separator was employed to collect and separate the fission fragments, while the second stage, combined with the following ZeroDegree spectrometer, served as a spectrometer for the particle identification of fragments. An achromatic aluminum energy degrader was used in the first stage of the BigRIPS separator to select the range of isotopes to be measured. Another energy degrader was used in the second stage in the case of the G3 setting, because further purification was needed. The angular acceptance of the BigRIPS separator was ±40 mrad horizontally and ±50 mrad vertically, while the momentum acceptance was set to ±3 % by using slits at the dispersive focus located at the mid-point of the first stage. The particle identification was performed based on the $\Delta E$-$TOF$-$B\rho$ method, in which the energy loss ($\Delta E$), time of flight ($TOF$) and magnetic rigidity (Bρ) were measured and used to deduce the atomic number ($Z$) and the mass-to-charge ratio ($A/Q$) of fragments. The $TOF$ and $B\rho$ values were measured in the second stage of BigRIPS separator, while the $\Delta E$ measurement was performed at the focal plane of ZeroDegree spectrometer. Note that in the search for new isotopes the $\Delta E$ measured at the end of the BigRIPS separator was used for the G2 and G3 settings. See Ref. [32] for details about the detectors used.

The identified fragments were implanted into the aluminum stopper which was located in air at the focal plane of ZeroDegree spectrometer. Prior to stopping, the ions were slowed down in an energy absorber consisting of a stack of aluminum plates. We chose the stopper thickness of 30 mm for the G1 setting, 10 mm for the G2 setting and 10 mm for the G3 setting, based on the stopping-range calculation of ATIMA [33]. The thickness of the absorber stack was 15 mm, 13 mm and 8 mm, respectively. The stopper was tilted by 26.6 degrees both horizontally and vertically with respect to the beam axis. This allowed reduction of the overall $\gamma$-ray attenuation in the stopper, because we could use a thinner stopper plate as well as avoid large attenuation in some directions. The stopper had an effective area of 90 × 90 mm$^2$. The beam profile on the stopper was monitored by using two sets of position-sensitive parallel plate avalanche counters (PPAC) [34]. The flight path between the production target and the stopper was 127.3 m long, and the flight time was typically 600 to 700 ns: *e.g.* 620 ns for $^{78}$Zn in the G1 setting, 630 ns for $^{108}$Zr in the G2 setting and 670 ns for $^{134}$Sn in the G3 setting.

The $\gamma$ rays emitted from the stopped ions were detected using three clover-type high-purity germanium (Ge) detectors that were placed approximately 7 cm on the left, right and lower sides of the stopper. The Ge detectors, each of which consists of four crystals, were operated in an add-back mode to construct $\gamma$-ray energy spectra. Lead collimators, 50 mm thick, were placed upstream of the Ge crystals to protect them from charged particles. The $\gamma$-ray detection was made based on the delayed $\gamma$ coincidence technique. We recorded the time interval between the $\gamma$-ray emission and the ion implantation by using a multi-hit time-to-digital convertor (TDC) module, and the obtained time spectrum was used to deduce the half-life of isomeric state. The range of the multi-hit TDC was set to 20 $\mu$s, and the time of implantation was taken from a plastic scintillation counter located at F3 in the BigRIPS separator [32]. We also recorded $\gamma\gamma$ coincidences among the Ge detectors. The time spectrum of $\gamma\gamma$ coincidence was constructed from the difference of the time intervals between detector hits, in order to identify true coincidence events. We investigated the $\gamma\gamma$ coincidences among all twelve individual crystals as well as among the three clover-type Ge detectors (operated in an add back mode). The former provided $\gamma\gamma$ coincidence data for low energy $\gamma$-rays.

The energy threshold and dynamic range of the Ge detectors were set to 50 and 4000 keV, respectively. The energy calibration and the measurement of absolute photo-peak efficiency were made by using standard $\gamma$ sources that were affixed on the front of the stopper. The measured overall efficiency of the three Ge detectors was 8.4 % for 122-keV $\gamma$ rays and 2.3 % for 1408-keV $\gamma$ rays in the case of the 30-mm thick aluminum stopper (G1 setting), and 11.9 % and 2.7 % in the case of the 10-mm thick aluminum stopper (G2 and G3 settings), respectively. The energy resolution (full width at half maximum) was 2.1 keV for 1-MeV $\gamma$ rays. The online detection efficiency for isomeric $\gamma$-rays was found to decrease to 70 % of the above offline values in the worst case. This is due to pileups caused by prompt $\gamma$- and X-rays, which are emitted prior to the isomeric decay in the process of the ion implantation. In the offline analysis, we excluded $\gamma$-ray events with the time interval approximately five times larger than the half-life to reduce $\gamma$-ray backgrounds due to accidental coincidences between the $\gamma$-ray emission and the ion implantation. We also excluded those with the time interval less than 200 ns to reduce background events caused by the prompt $\gamma$- and X-rays. In investigating the $\gamma\gamma$ coincidences, the prompt events were rejected by gating on a two-dimensional plot of energy versus time.

Figures 1(a) and 1(b) show the $Z$ versus $A/Q$ particle identification (PID) plots obtained in the G2 setting, corresponding to PID plots without (a) and with (b) the delayed $\gamma$ coincidence. Figure 1(a) shows clear identification of fission fragments including their charge states. Thanks to the excellent $A/Q$ resolution, the peaks for fully-stripped ($Q = Z$) and hydrogen-like ($Q = Z$-1) ions are well separated from each other for each isotopic line. Figures 1(b) shows the observation of isomers in which isotope events having isomeric decays are clearly enhanced. Strongly observed isomers are labeled in an enlarged PID plot shown in Fig. 1 (c). Here we selected $\gamma$-ray events with the time interval between 0.2 and 5 $\mu$s to exclude the accidental coincidence events as well as the prompt events. The probability of the accidental coincidence was estimated from high implantation statistics isotopes that do not contain isomeric decays. The estimate is that on average only 2.5 % of implants would generate a background coincidence.

A delayed $\gamma$-ray energy spectrum has been obtained for each isotope by applying the time gate as well as a PID gate, and we have found a number of $\gamma$-ray peaks that we attribute to isomeric decays. If the statistics of the $\gamma$-ray peaks were poor, we made a significance test [35] to confidently identify them against the continuous background $\gamma$-rays that were generated by accidental coincidences. The statistical probability of misidentification called the 'p-value' [35] was evaluated based on the background spectrum obtained from high implantation statistics isotopes that do not contain isomeric decays. The isomeric $\gamma$-rays shown in the next chapter were those identified with a p-value smaller than 0.1 %, which corresponds to a confidence level better than 99.9 %.

Background $\gamma$-ray peaks were observed in the delayed $\gamma$-ray energy spectra. We carefully analyzed those $\gamma$-ray peaks that appeared in multiple spectra, in order to find the origin of background $\gamma$-rays. In addition to natural background $\gamma$-rays such as $^{40}$K, the following three origins were identified: i) neutron-induced reactions such as $^{72}$Ge(n,n')$^{72}$Ge(691.55 keV, $0^+$, 444.2 ns) and $^{19}$F(n,n'$\gamma$)$^{19}$F(197.14 keV, $5/2^+$, 89.3 ns) [36] that occur around the $\gamma$-ray detectors, ii) $\beta$ activities such as $^{24}$Na and $^{26}$Al produced from the stopper material in the process of the ion implantation and iii) isomeric $\gamma$-decays from neutron-rich nuclei produced by secondary reactions of the implanted fragments.

The total number of the implanted fragments was 5.78 x $10^7$, 4.41 x $10^7$ and 8.46 x $10^7$ in the G1, G2 and G3 settings, respectively, and the corresponding implantation rate is 529, 270 and 871 pps on average. The total number of the implanted fragments having isomeric decays was 8.97 x $10^6$, 1.03 x $10^7$ and 3.33 x $10^7$ in the G1, G2 and G3 settings, respectively, and the corresponding implantation rate is 82.1, 63.1 and 343 pps on average. The net observation time in the G1, G2 and G3 settings was 30.3 h, 45.3 h and 27.0 h, respectively. The counting rates of the isomeric $\gamma$-rays, including Compton scattered events, are determined by the isomer ratio, the decay loss in flight, the internal conversion coefficients, and the $\gamma$-ray detection efficiency. The isomer ratio varies with the observed isomers, ranging from several percent to a few tens of percent.

## III. RESULTS AND DISCUSSION

The microsecond isomers that we have observed in the present experiment are summarized in Tables I and II along with the BigRIPS settings, the number of implanted isotopes, the energy of isomeric $\gamma$-rays ($E_\gamma$) and the half-lives ($T_{1/2}$). The tables list 54 isomers in total, including the new isomers listed in Table I. The $\gamma$-ray peaks were fit using the code gf3 in the RADWARE software package [37]. The systematic uncertainty of $\gamma$-ray energies was estimated to be ±0.5 keV from the accuracy of the energy calibration with standard $\gamma$ sources. The statistical uncertainty of $\gamma$-ray energies was much smaller than the systematic one in most cases. The half-lives were deduced from fitting analysis of the time spectra of isomeric $\gamma$-decays, in which the maximum likelihood method was used assuming exponential decay component(s) plus a constant background. In cases where more than one $\gamma$-decay from a given isotope showed the same half-life, we added the time spectra to improve the half-life determination. The half-life values given in Tables I and II are those obtained from the latter time spectra. Some more details of the fitting analysis are given in their footnotes. The new isomers discovered in the present experiment were: $^{59}Ti^{m}$, $^{90}As^{m}$, $^{92}Se^{m}$, $^{93}Se^{m}$, $^{94}Br^{m}$, $^{95}Br^{m}$, $^{96}Br^{m}$, $^{97}Rb^{m}$, $^{108}Nb^{m}$, $^{109}Mo^{m}$, $^{117}Ru^{m}$, $^{119}Ru^{m}$, $^{120}Rh^{m}$, $^{122}Rh^{m}$, $^{121}Pd^{m}$, $^{124}Pd^{m}$, $^{124}Ag^{m}$ and $^{126}Ag^{m}$. Figures 2 and 3 show their energy spectra and time spectra, respectively. Half-lives have been determined for these new isomers for the first time, except for $^{124}Pd^{m}$ and $^{126}Ag^{m}$ which were observed with $T_{1/2}$ > 20 $\mu$s. We classify $^{59}Ti^{m}$, $^{117}Ru^{m}$, $^{120}Rh^{m}$, $^{121}Pd^{m}$ and $^{124}Ag^{m}$ as new isomers due to incomplete or differing results in previous reports. In the cases of $^{117}Ru^{m}$, $^{121}Pd^{m}$ and $^{124}Ag^{m}$, only the $\gamma$-ray energy spectra were presented without any analysis in a conference proceedings report [38] and details have not been published in a refereed paper. In the case of $^{120}Rh^{m}$, the $\gamma$-ray energy spectrum published in Ref. [39] differs substantially from what was observed in the present work. For $^{59}Ti^{m}$ the $\gamma$-ray energy given in the previous report [40] differs significantly with our present data by as much as 8 keV.  Figures 4 and 5 typically show the $\gamma\gamma$ coincidence spectra of $^{108}Nb^{m}$ and $^{119}Ru^{m}$, respectively. The $^{108}Nb^{m}$ spectra represent the cases with good statistics and the $^{119}Ru^{m}$ spectra those with typical statistics.

The observed isomers are shown in the nuclear chart of Fig. 6. We identified a number of new and known isomers in neutron-rich exotic nuclei between $N$ ~ 50 and 82 from $^{78}Ni$ to $^{132}Sn$. Isomeric states with $N$ ~ 58 to 61 have been known to be generated due to shape coexistence of spherical and prolate-deformed states [6]. In the present work several new isomers with $Z$ < 38 were discovered, allowing us to investigate how shape coexistence evolves over the entire region. Several isomers were also observed in the region of $N$ ~ 65 to 70, in which a variety of nuclear shapes, including prolate, oblate, triaxial and tetrahedral shapes [11, 41-43], have been suggested. Furthermore we made the first broad exploration for the $N$ ~75 region and discovered several new isomers. In addition to these observations, isomers in the lighter-mass region were also explored, some of which are of importance to study the shell evolution in the neutron-rich exotic nuclei between $N$ ~ 28 and 50 [44-46].

Table I lists the measured $\gamma$-ray relative intensities ($I_\gamma$) for the new isomers. Those of known isomers are listed in Table III. In deducing the relative $\gamma$-ray intensities, we used the $\gamma$-ray detection efficiency that was evaluated by numerical simulation using the Monte Carlo code GEANT3 [47]. This allowed us to realistically take into account the $\gamma$-ray attenuation using the detailed geometrical layout of the stopper and detectors. The simulation well reproduced the measured $\gamma$-ray detection efficiency using standard $\gamma$ sources that were affixed to the front surface of the stopper. We made the simulation for isomeric $\gamma$-rays incorporating the realistic beam-spot profile and stopping-range distributions in the stopper, which were obtained based on the measured positions and $B\rho$ distributions of implanted fragments. The previously reported $\gamma$-ray relative intensities, such as those of $^{78}$Zn$^m$ (G1), $^{95}$Kr$^m$ (G1 and G2), $^{100}$Sr$^m$ (G2), $^{127}$Cd$^m$ (G2 and G3), $^{128}$Cd$^m$ (G2 and G3), $^{129}$In$^m$ (G3), $^{131}$Sn$^m$ (G3), $^{132}$Sn$^m$ (G3) and $^{134}$Sn$^m$ (G3), are reproduced using the $\gamma$-ray detection efficiency thus obtained.

We have also obtained some new results on the known isomers. They are indicated in Table II by boldface with an underline. New $\gamma$-ray peaks were identified for $^{92}$Br$^m$, $^{108}$Zr$^m$ and $^{125}$Ag$^m$. The half-lives of $^{82}$Ga$^m$ and $^{92}$Br$^m$ were determined for the first time. The precision of the half-lives was significantly improved for the following isomers: $^{95}$Kr$^m$, $^{98}$Rb$^m$, $^{108}$Zr$^m$, $^{109}$Nb$^m$, $^{113}$Tc$^m$, $^{125}$Ag$^m$ and $^{136}$Sb$^m$. Table II shows the comparison with the previous measurements. In most cases our half-life measurements are in agreement with those given in the literature.

The present experimental data allowed us to propose level schemes for some of the observed isomers. Level schemes were constructed based on the $\gamma$-ray relative intensities, the $\gamma\gamma$ coincidence information and the correspondence of decay curves as well as the $\gamma$-ray energies and the half-lives. The energy sum, the cascade and side feeding, the intensity balance, and the multi-polarity and reduced transition probability for isomeric $\gamma$-ray transitions were considered as well. Systematics of the level schemes and nuclear isomerism in neighboring nuclei as well as their theoretical studies were also compared, if available. In case of odd-odd nuclei, we relied on the so-called Nordheim rule [48], in those cases where the configuration of single particle orbitals could be inferred from the systematics of neighboring nuclei.

If the $\gamma$-ray energies were relatively low, the intensity balance analysis of cascading $\gamma$-rays allowed us to deduce the multi-polarities of transitions since the intensity correction for internal conversion is sensitive to the multi-polarity [4, 23, 49]. This could not be applied in cases near the energy threshold where the $\gamma$-ray detection efficiency had the large uncertainty. The intensity correction was made using the total internal conversion coefficients calculated with the code BrIcc [49]. Furthermore the so-called recommended upper limits (RUL) analysis [50, 51] was performed for the reduced transition probability to determine the upper limit of multi-polarity for isomeric transitions.

Figures 7, 10, 11 and 15-18 show the level schemes that we propose for the new isomers: $^{59}$Ti$^{m}$, $^{94}$Br$^{m}$, $^{95}$Br$^{m}$, $^{97}$Rb$^{m}$, $^{108}$Nb$^{m}$, $^{109}$Mo$^{m}$, $^{117}$Ru$^{m}$, $^{119}$Ru$^{m}$, $^{120}$Rh$^{m}$, $^{122}$Rh$^{m}$, $^{121}$Pd$^{m}$ and $^{124}$Ag$^{m}$ and for the known isomers: $^{82}$Ga$^{m}$, $^{92}$Br$^{m}$ and $^{98}$Rb$^{m}$. These level schemes have been constructed for the first time. Figures 14 and 20 show our proposed revisions for the previously reported level schemes of $^{108}$Zr$^{m}$ and $^{125}$Ag$^{m}$. We could infer the nature of the nuclear isomerism for some isomers. Detailed discussions of the proposed level schemes are given in the following sections.

## A. $^{59}$Ti$^{m}$

The proposed level scheme for $^{59}$Ti$^{m}$ is shown in Fig. 7. A new isomeric $\gamma$-ray at 109.0 keV with a half-life of $0.587^{+0.057}_{-0.051}$ $\mu$s is observed. Based on the systematics in neighboring nuclei, we propose that the observed isomer in $^{59}$Ti is an $E2$ isomer with spherical structure, which is generated due to a small transition energy. This type of isomer often appears in spherical nuclei. The reduced $E2$ transition probability $B(E2)$ of the isomeric transition is $3.68^{+0.37}_{-0.34}$ W.u., consistent with spherical structure. The possibility of higher multi-polarities is excluded, because the reduced transition probability significantly exceeds the recommended upper limits (RUL). The possibility of hindered $E1$ or $M1$ transition is also excluded, because such an isomer, which usually appears as a shape isomer, has not been observed in the neighboring isotopes. Furthermore similar $E2$ isomers such as $^{54}$Sc$^{m}$ [20, 44] and $^{56}$Sc$^{m}$ [44] appear in the vicinity of $^{59}$Ti$^{m}$, supporting our proposed origin.

The spin and parity ($J^{\pi}$) of the ground state in $^{59}$Ti is tentatively assigned as $5/2^{-}$, based on the systematics of $^{55}$Ti [52] and $^{57}$Ti [44, 53], which indicates that the $\nu f_{5/2}$ single-particle orbital is located above the $\nu p_{1/2}$ orbital in these neutron-rich Ti isotopes. Assuming such a shell ordering and similarity to $^{57}$Ti, the $J^{\pi}$ of the 109.0-keV isomeric state in $^{59}$Ti is tentatively assigned as $1/2^{-}$, with a configuration dominated by the $\nu p_{1/2}$ hole state. These assignments are consistent with the proposed $E2$ multi-polarity for the isomeric transition. According to Ref. [44], the $1/2^-$ state in $^{57}$Ti is located at 364 keV above the $5/2^-$ ground state. This implies that the sub-shell gap between the $\nu p_{1/2}$ and $\nu f_{5/2}$ orbitals gets smaller as the neutron number increases, thus generating the isomeric transition in $^{59}$Ti.

### B. $^{82}$Ga$^{m}$

The proposed level scheme for $^{82}$Ga$^{m}$ is shown in Fig. 7 and the $\gamma$-ray energy spectrum and decay curve are shown in Fig. 8. A single isomeric $\gamma$-ray at 141.0 keV is observed, confirming the previous measurement in Ref. [27] in which a $\gamma$-ray at 141 keV was observed. The 197-keV peak is attributed to $^{19}$F(n,n'$\gamma$)$^{19}$F. The half-life is measured to be $0.098^{+0.010}_{-0.009}$ $\mu$s for the first time. We propose that the observed isomer in $^{82}$Ga is an $E2$ isomer with spherical structure, based on the systematics in neighboring nuclei. In this case the deduced $B(E2)$ value is $4.09^{+0.43}_{-0.39}$ W.u., consistent with our interpretation. The possibility of other multi-polarities is excluded similarly to the case of $^{59}$Ti$^{m}$. The ground state and the 141.0-keV isomeric state in $^{82}$Ga$^{m}$ are tentatively assigned as $J^{\pi} = 0^-$ and $2^-$, respectively, according to the shell ordering in this region and the Nordheim rule for odd-odd spherical nuclei [48].

According to the $\beta$-decay spectroscopy of $^{83}$Ge [45] and $^{81}$Ga [46], it can be assumed that the valence proton and neutron of the ground state of $^{82}$Ga occupy the $\pi f_{5/2}$ and $\nu d_{5/2}$ orbitals, respectively. In such a configuration the Nordheim rule predicts that the $0^-$ coupling is energetically favorable for the ground state. Accordingly the isomeric state in $^{82}$Ga has been assigned as $J^{\pi} = 2^-$. We propose that the configuration of the isomeric state is dominated by $[\pi f_{5/2} \otimes \nu s_{1/2}]2^-$, based on the ordering and spacing of the single-particle orbitals in the relevant neutron-rich region [45, 46]. Furthermore it has been pointed out [45] that the $\nu s_{1/2}$ orbital in the odd-mass $N = 51$ isotones systematically goes down as the proton number decreases. The first $1/2^+$ state is located at 1031 keV for $^{89}$Sr, 532 keV for $^{87}$Kr, 462 keV for $^{85}$Se and 260keV for $^{83}$Ge. The present data suggest that the $\nu s_{1/2}$ orbital goes down further in $^{82}$Ga, resulting in the occurrence of the $E2$ isomer in $^{82}$Ga. Based on our observation, we cannot rule out the possibility that the $\nu s_{1/2}$ orbital is lower than the $\nu d_{5/2}$ orbital, in which case the $J^{\pi}$ assignments to the ground and isomeric states would be reversed.

### C. $^{90}$As$^{m}$, $^{92}$Se$^{m}$, $^{93}$Se$^{m}$ and $^{96}$Br$^{m}$

There is not enough information to propose level schemes for these new isomers, as we cannot rule out missing $\gamma$ transitions due to low statistics. However, the discovery of these isomers provides us with future opportunities to investigate how the nuclear shape and shape coexistence evolve in the region of $N$ ~ 58 to 61 [6] with decreasing Z number. Discussions on the evolution of nuclear shape in other neutron-rich nuclei in this region are given in the succeeding sections.

### D. $^{92}$Br$^{m}$

The $\gamma$-ray energy and time spectra of $^{92}$Br$^{m}$ are shown in Fig. 9. A total of eleven isomeric $\gamma$-rays are observed, including two new $\gamma$-rays at 197.6 and 898.3 keV. The half-life of $^{92}$Br$^{m}$ is determined for the first time. We find some disagreement with the previous measurement [27] as to the observed $\gamma$-ray peaks. Nine of the previously reported $\gamma$-rays are confirmed, but those at 169 and 295 keV [27] are not identified in the present experiment. The level scheme of $^{92}$Br$^{m}$ was not proposed in Ref. [27], because they were unable to construct a plausible scheme using energy sum and intensity relations. In this previous experiment $\gamma\gamma$ coincidence was not measured.

Based on the energy sum analysis for the observed $\gamma$-rays, we propose the level scheme shown in Fig. 10 for $^{92}$Br$^{m}$, in which two isometric states with similar half-lives are locate at 662 and 1138 keV. The energy sum of the 197.6- and 259.5-keV transitions agrees well with that of the 154.9- and 301.9-keV transitions. The following energy sum relations are also found: 239.5 keV + 898.3 keV ≈ 98.6 keV + 1039.8 keV ≈ 98.6 keV + 259.5 keV + 778.8 keV ≈ 1138 keV, 98.6 keV + 139.7 keV = 238.3 keV ≈ 239.5 keV, and 259.5 keV + 778.8 keV = 1038.3 keV ≈ 1039.8 keV. According to these relations, the observed $\gamma$-rays are uniquely assigned to the level scheme. Furthermore, in the present experiment $\gamma\gamma$ coincidences are identified between the 98.6-keV $\gamma$-ray and those at 106.3, 154.9, 259.5 and 898.3 keV, between the 106.3-keV $\gamma$-ray and those at 154.9 and 259.5 keV, and between 259.5-keV $\gamma$-ray and those at 154.9, 197.6 and 778.8 keV, consistent with the proposed level scheme. The intensity balance of $\gamma$-rays seems fairly consistent as well. Note that we could not observe the 43-keV transition due to our energy threshold. The half-lives of the 662- and 1138-keV isomeric states are deduced to be $0.089^{+0.007}_{-0.008}$ and $0.084^{+0.010}_{-0.009}$ $\mu$s, respectively. Here we use the $\gamma$-ray peaks at 106.3, 154.9 and 301.9 keV for the 662-keV state and those at 778.8, 898.3 and 1039.8 keV for the 1138-keV state. The half-life of 662-keV state was deduced assuming two decay components plus a constant background, because the background events of the 106.3-, 154.9- and 301.9-keV $\gamma$-rays contained Compton-scattered events originating from both 662- and 1138-keV states. We deduced the decay component of these background events by fitting the neighboring events on the high-energy side of the relevant $\gamma$-ray peak. The obtained decay parameter was fixed in the fitting analysis to deduce the half-life of 662-keV state.

The RUL analysis restricts the multi-polarity of the isomeric transition from the 662-keV state to *E*1, *M*1 or *E*2. The reduced transition probability reveals hindrance if the multi-polarity is assigned as *E*1 or *M*1. The isomeric state at 1138 keV is likely due to high multi-polarities of the transitions such as *M*2 or *E*3. In the neighboring odd-odd $N$ = 57 isotone $^{94}$Rb, two isomeric states with half-lives similar to those of $^{92}$Br$^m$ were recently identified at 1485.2 and 2074.8 keV, and it was proposed that they were spherical isomers caused by fully-aligned configurations of the high-*j*, single-particle orbitals $\pi g_{9/2}$, $\nu g_{7/2}$ and $\nu h_{11/2}$ [54]. The ground state in $^{94}$Rb is characterized by a configuration of $[\pi f_{5/2}\otimes\nu s_{1/2}]3^-$ and the 1485.2- and 2074.8-keV isomeric states are suggested to have spherical configurations of $[\pi g_{9/2}\otimes\nu g_{7/2}]8^+$ and $[\pi g_{9/2}\otimes\nu h_{11/2}]10^-$, respectively [54]. We speculate by analogy that the ground state in $^{92}$Br have the same spherical configuration with $J^\pi = 3^-$ and the observed isomers in $^{92}$Br are likely caused by a similar mechanism. Further experimental and theoretical studies are needed to clarify the nature of these isomers conclusively. Moreover, in the present experiment, an isomeric transition are also observed in $^{90}$As for the first time, suggesting that similar isomerism may occur in this lighter $N$ = 57 isotone.

### E. $^{94}$Br$^m$

A total of six new isomeric $\gamma$-rays at 88.0, 92.1, 115.5, 165.7, 179.0 and 202.6 keV with a half-life of 0.530(15) $\mu$s are identified in the $\gamma$-ray energy spectrum of $^{94}$Br$^m$. Our proposed partial level scheme is shown in Fig. 11. A $\gamma\gamma$ coincidence is observed between the 92.1- and 202.6-keV $\gamma$-rays. A $\gamma\gamma$ coincidence between 115.5- and 179.0-keV $\gamma$-rays is also observed, although the statistics are low. The energy sums of the cascading $\gamma$-rays, which are 92.1 keV + 202.6 keV = 294.7 keV and 115.5 keV + 179.0 keV = 294.5 keV, agree well with each other. The 179.0-keV transition is placed in the upper part of the cascade because the 88.0-keV $\gamma$-ray is observed. Inverse ordering of the cascading $\gamma$-rays is also allowed, if we only rely on these results. The 165.7-keV $\gamma$-ray cannot be assigned to the level scheme, although its decay curve is consistent with the half-life of $^{94}$Br$^m$.

In the neighboring $N$ = 59 isotones, such as $^{95}$Kr [55], $^{96}$Rb [56], $^{97}$Sr [57, 58], $^{98}$Y [59] and $^{99}$Zr [60], the low-lying states systematically reveal spherical structure. *E*2 isomers with spherical structure were previously observed at the excitation energies below ~400 keV in the odd-odd isotone $^{98}$Y as well as the even-odd isotones $^{95}$Kr, $^{97}$Sr and $^{99}$Zr. The present experiment confirms these isomers in $^{95}$Kr and $^{97}$Sr. We propose that the 295-keV state in $^{94}$Br$^m$ is an isomer with spherical structure, based on the systematics in these neighboring $N$ = 59 isotones. The intensity balance analysis for the cascading 92.1- and 202.6-keV $\gamma$-rays in $^{94}$Br$^m$ allows us to assign the multi-polarity of 92.1-keV transition to be $E$2, because it gives the most consistent intensity balance after the correction of internal conversion. Here the multi-polarity of the 202.6-keV $\gamma$-ray is assumed to be $E$1, $M$1 or $E$2, because the higher multi-polarities would give a longer half-life for the relevant state. Hence, taking into account the branching of $\gamma$ rays, the $B$($E$2) value of the 92.1-keV transition is estimated to be 2.49(24) to 2.45(24) W.u., consistent with our proposal for $^{94}$Br$^m$. Here multi-polarities of $E$1 to $M$2 are considered for the 179.0-keV transition, based on the RUL. We speculate that the ground state in $^{94}$Br has a spherical configuration of $[\pi f_{5/2}\otimes\nu s_{1/2}]2^-$ in analogy to the neighboring odd-odd $N$ = 59 isotone $^{96}$Rb [56]. Furthermore, in the present experiment, isomeric $\gamma$-rays in $^{93}$Se are also observed for the first time, implying the existence of similar spherical isomer in this lighter $N$ = 59 isotone.

In the $N$ = 59 isotones $^{96}$Rb [56], $^{97}$Sr [57, 58], $^{98}$Y [59] and $^{99}$Zr [60], prolate-deformed states were systematically identified above the low-lying spherical states, revealing shape coexistence, and a band head of the deformed states, such as the [404]9/2$^+$ states in $^{97}$Sr [57, 58] and $^{99}$Zr [60], exhibits isomeric transitions. These deformed states have not been previously identified in $^{95}$Kr [55]. In the present experiment we do not observe any isomeric $\gamma$-rays from such deformed states in $^{95}$Kr and $^{94}$Br.

### F. $^{95}$Br$^m$

The proposed level scheme for $^{95}$Br$^m$ is shown in Fig. 11. A single isomeric $\gamma$-ray at 537.9 keV with a half-life of $6.67^{+1.10}_{-0.85}$ $\mu$s is observed for the first time. We propose that the 537.9-keV state in $^{95}$Br is a prolate-deformed shape isomer located above the spherical ground state, based on the systematics of shape coexistence in the $N$ = 60 isotones. These systematics will be discussed with $^{97}$Rb$^m$ in the next section. The reduced transition probability of the isomeric transition reveals very hindered values up to the $M$2 multi-polarity, consistent with our proposed isomerism. The upper limit of possible multi-polarity is $E$3, according to the RUL analysis. The $N$ = 60 systematics allows us to tentatively assign the $J^\pi$ of the ground state as 5/2$^-$ and that of the 537.9-keV isomeric state as [312]3/2$^-$, [310]1/2$^-$ or [440]1/2$^+$. The details of $J^\pi$ assignments are also discussed in the next section.

## G. $^{97}$Rb$^m$

The proposed level scheme for $^{97}$Rb$^m$ is shown in Fig. 11. A single isomeric $\gamma$-ray at 77.1 keV with a half-life of $6.33^{+0.37}_{-0.34}$ $\mu$s is observed for the first time. The ground state of $^{97}$Rb is known to exhibit a large static prolate deformation ($\varepsilon = 0.29$) characterized by the Nilsson orbit [431]3/2$^+$ from spin and magnetic moments determined by laser spectroscopy [61]. The deformation is supported by the mass systematics in the relevant neutron-rich region [62] (see Fig. 1 in the reference). We propose that the 77.1-keV state in $^{97}$Rb is a shape isomer, which has a spherical shape and decays to the deformed ground state, based on the systematics of shape coexistence in neighboring nuclei.

It is known that a sudden onset of prolate deformation occurs in the ground states of $N$ = 60 isotones heavier than $^{97}$Rb, which was observed as dramatic changes in the isotope shift and the two-neutron separation energy [62]. It is also known that shape coexistence occurs in the even-even $N$ = 60 isotones, in which a spherical second 0$^+$ state systematically appears above a prolate-deformed ground state at an excitation energy below 1 MeV [6]. Furthermore the spherical 0$^+$ state in these nuclei goes down with decreasing proton number [6]: *e.g.* 698 keV for $^{102}$Mo [63], 331 keV for $^{100}$Zr [8], 215 keV for $^{98}$Sr [7], and the ground state of $^{96}$Kr does not exhibit such deformation according to the mass systematics [62] and spectroscopic studies [64]. This systematic behavior of $N$ = 60 nuclei imply that the energy levels of spherical and deformed states get closer as the proton number decreases and reverses between $^{98}$Sr and $^{96}$Kr. We speculate that the $^{97}$Rb isotope is a transitional nucleus in which the spherical and deformed structures energetically compete with each other at a low excitation energy. It should be noted that similar shape coexistence was suggested in the odd $N$ = 60 isotone $^{99}$Y, in which a spherical state is located at 599 keV above the deformed ground state [65]. We tentatively assign the $J^\pi$ of the 77.1-keV state in $^{97}$Rb as 5/2$^-$, based on the $J^\pi$ assignment of a spherical ground state in $^{95}$Rb [66]. Accordingly the multi-polarity of the 77.1-keV transition has been assigned as $E1$, which reveals a very hindered $B(E1)$ value, consistent with our proposed isomerism. A little information is available for excited states in $^{96}$Kr. Their investigations, including search for deformed excited states in $^{96}$Kr, are awaited to further clarify the evolution of shape coexistence in the $N$ = 60 isotones. In the present measurement we did not observed any isomeric transitions in $^{96}$Kr although the statistics were as high as those of $^{95}$Kr$^m$.

Based on the above discussion of $N = 60$ systematics including $^{97}Rb^m$, we propose the level scheme shown in Fig. 11 for $^{95}Br^m$, with the spherical and deformed states reversed. The $J^\pi$ assignment to the deformed state at 538 keV is made based on the Nilsson diagram in Fig. 7 of Ref. [67], assuming that the state has the same prolate-deformation parameter as $^{97}Rb$ ($\varepsilon = 0.29$).

### H. $^{98}Rb^m$

The proposed level scheme for $^{98}Rb^m$ is shown in Fig. 11 and the $\gamma$-ray energy spectrum and decay curve are shown in Fig. 12. Three isomeric $\gamma$-rays at 116, 124 and 178 keV with a half-life of $700^{+60}_{-50}$ ns were previously reported [27] for $^{98}Rb^m$. In the present work only two isomeric $\gamma$-rays at 123.7 and 178.4 keV with a half-life of 0.358(7) $\mu$s are observed. The 116-keV $\gamma$-ray is not confirmed, even though our photo-peak counts of the 123.7-keV $\gamma$-ray are approximately ten times more than those reported in Ref. [27]. The 116-keV $\gamma$-ray should have been observed in this experiment if it originates from the same isomeric state. We propose the level scheme shown in Fig. 11 for $^{98}Rb^m$, because a $\gamma\gamma$ coincidence is not observed between the 123.7- and 178.4-keV $\gamma$-rays. The statistics of the present measurement can rule out that the $\gamma$-rays are in a cascade.

In the neighboring $N = 61$ isotones, such as $^{99}Sr$ [68], $^{100}Y$ [69], $^{101}Zr$ [70] and $^{102}Nb$ [69], the low-lying states systematically reveal prolate-deformation. Based on the RUL analysis for reduced transition probability, we propose that the 178.4-keV state in $^{98}Rb$ is likely a shape isomer. The RUL analysis considering the branching of transitions restricts the multi-polarities of the 178.4-keV transition and the unobserved 54.7-keV transition to $E1$, $M1$ or $E2$. The reduced transition probability reveals hindered values for all the possible multi-polarities for the 178.4-keV transition, supporting our proposed isomerism. In the case of the unobserved 54.7-keV transition, the transition probability would have a significant hindrance except for an $E2$ multi-polarity. The present data cannot rule out that the ordering of the unobserved 54.7-keV transition and the 123.7-keV transition could be reversed. However this does not affect the above discussion. In the present experiment, an isomeric $\gamma$-ray is also observed in $^{96}Br$ for the first time, implying that similar isomerism occurs in this lighter $N = 61$ isotone. Further investigations are needed to clarify the nuclear isomerism in the $N = 61$ isotones conclusively.

### I. $^{108}Zr^m$

The $\gamma$-ray energy and time spectra of $^{108}Zr^m$ are shown in Fig. 13 and our proposed level scheme for $^{108}Zr^m$ is shown in Fig. 14. A total of thirteen isomeric $\gamma$-rays at 174.4, 277.9, 347.2, 364.2, 426.3, 432.3, 478.3, 485.3, 603.9, 642.3, 773.6, 795.9 and 828.3 keV with a half-life of $0.536^{+0.026}_{-0.025}$ $\mu$s are observed in this isomer. Among them the observation of five $\gamma$-rays at 174.4, 277.9, 347.2, 478.3 and 603.9 keV was previously reported along with the half-life value of 620(150) ns [43]. The authors of Ref. [43] proposed a level scheme of rotational-like band up to $4^+$ consisting of the 174.4- and 347.2-keV transitions, based on the systematics of the $2^+_1$ and $4^+_1$ states in the lighter even-even Zr isotopes, although a $\gamma\gamma$ coincidence was not observed. In the present experiment we observe a $\gamma\gamma$ coincidence between the 174.4- and 347.2-keV $\gamma$-rays, supporting their proposed level scheme.

We extended the level scheme based on the energy and intensity relation as well as the $\gamma\gamma$ coincidence information. All the observed $\gamma$-rays are uniquely assigned to the level scheme in Fig. 14, including the ground-state rotational band which is extended up to the $8^+$ state. The energy sum analysis strongly supports the proposed level scheme, although there is some inconsistency in the relative intensity of the 426.3-keV $\gamma$-ray. We tentatively assign the isomeric state of $^{108}Zr$ to the level at 2075 keV. The recent theoretical calculation [71] predicts that a two-quasineutron state with $J^\pi = 6^+$ appears at 1.997 MeV in $^{108}Zr$. Our proposed isomeric state at 2075 keV is located close to this predicted high-$K$ state, implying that the isomeric state in $^{108}Zr$ could be a $K$ isomer. More details will be presented in our separate paper.

### J. $^{108}Nb^m$

The proposed level scheme for $^{108}Nb^m$ is shown in Fig. 15. Four new isomeric $\gamma$-rays at 64.3, 77.6, 89.0 and 102.2 keV with a half-life of 0.109(2) $\mu$s are observed. We clearly observe a $\gamma\gamma$ coincidence between the 77.6- and 89.0-keV $\gamma$-rays as shown in Fig. 4. A $\gamma\gamma$ coincidence is also observed between the 64.3- and 102.2-keV $\gamma$-rays. The energy sums of these cascading $\gamma$-rays, which are 77.6 keV + 89.0 keV = 166.6 keV and 64.3 keV +102.2 keV = 166.5 keV, agree well with each other. Furthermore a $\gamma\gamma$ coincidence between the 64.3- and 89.0-keV $\gamma$-rays is observed. If we only rely on these results, reversed ordering of the cascading $\gamma$-rays is also possible.

The ground state of $^{108}Nb$ is suggested to be a prolate-deformed state with $J^\pi = 2^+$ and characterized by a configuration of $\pi[422]5/2^+\nu[411]1/2^+$, according to the recent $\beta$-decay spectroscopy of $^{108}Nb$ [72]. We propose to assign the multi-polarities of the cascading 77.6-keV and 89.0-keV $\gamma$-rays as $E1$ or $M1$, and $E2$, respectively, because such assignments give the most consistent intensity balance. If the ordering of these two $\gamma$-rays is that shown in Fig.10, the 89.0-keV isomeric $E2$ transition has a normal value in the Weisskopf estimate, suggesting that the 167-keV isomeric state in $^{108}$Nb is due to the small transition-energy between the deformed states in different bands.

If the ordering is reversed, the 77.6-keV isomeric transition becomes a hindered $E1$ or $M1$. In this case we propose that the 167-keV is an oblate-deformed shape isomer which decays to the prolate-deformed states. In fact such shape coexistence has been recently suggested in the neighboring isomer $^{109}$Nb$^{m}$ [42]. Here the transition probability of the 89.0-keV $E2$ transition should have an enhanced value, because otherwise the corresponding state would have a long half-life. Further experimental studies are needed to propose the level scheme and the nuclear isomerism conclusively. The branching of the isomeric transitions was taken into account in deducing the transition probability.

### K. $^{109}$Nb$^{m}$, $^{112}$Tc$^{m}$ and $^{113}$Tc$^{m}$

It was recently suggested in Ref. [42] that the isomer $^{109}$Nb$^{m}$ is an oblate-deformed shape isomer located above prolate-deformed states. In the present experiment three isomeric $\gamma$-rays at 117.0-, 196.2- and 312.2-keV are observed, confirming all the previously reported $\gamma$-rays in $^{109}$Nb$^{m}$ [42]. The present measurement has approximately ten times more statistics than the previous one, so that the precision of the half-life is improved significantly. Furthermore the improved statistics allow us to observe a $\gamma\gamma$ coincidence between 117.0- and 196.2-keV transitions for the first time, supporting the level scheme proposed in Ref. [42]. In the case of $^{112}$Tc$^{m}$ and $^{113}$Tc$^{m}$, we confirm the previously reported isomeric $\gamma$-rays: the 92- and 258-keV $\gamma$-rays in $^{112}$Tc$^{m}$ [41] and the 114-keV $\gamma$-ray in $^{113}$Tc$^{m}$ [41]. The isomeric states in $^{112}$Tc$^{m}$ and $^{113}$Tc$^{m}$ are suggested to be a triaxial-deformed shape isomer located above oblate-deformed states [41].

### L. $^{109}$Mo$^{m}$

The proposed level scheme for $^{109}$Mo$^{m}$ is shown in Fig. 15. A single $\gamma$-ray peak at 69.7 keV with a half-life of $0.194^{+0.076}_{-0.049}$ $\mu$s is observed for the first time. The level scheme of $^{109}$Mo was previously investigated by measuring prompt $\gamma$-rays following spontaneous fission of $^{248}$Cm [73]. In that experiment the ground state rotational band and the rotational band built on the 222-keV state were identified for $^{109}$Mo, and the ground state of $^{109}$Mo was suggested to be prolate-deformed and characterized by the Nilsson orbit [402]5/2$^{+}$ [73].

The RUL analysis limits the multi-polarity of the isomeric transition up to *E*2. We propose that the isomeric state in $^{109}$Mo is an *E*2 isomer whose transition takes place between the band heads of different deformed bands, in analogy to the neighboring isomer $^{107}$Mo$^{m}$ in which a similar isomeric transition was observed between the 65.4-keV state and the ground state [11]. According to Ref. [11], the ground state and the 65.4-keV state in $^{107}$Mo$^{m}$ are the band heads of the 5/2$^{+}$ rotational band and the 1/2$^{+}$ rotational band, respectively, and the 65.4-keV transition is an *E*2 transition with *B*(*E*2) = 5.9(5) W.u.. The configurations of these states are characterized by [413]5/2$^{+}$ and [411]1/2$^{+}$, respectively. Furthermore these two bands were suggested to have similar prolate and triaxial deformation parameters [11]. In the case of $^{109}$Mo$^{m}$ the deduced *B*(*E*2) value is $10.5^{+4.2}_{-2.8}$ W.u., which is similar to that of $^{107}$Mo$^{m}$, supporting our proposed isomerism. We tentatively assign the $J^{\pi}$ of the isomeric state in $^{109}$Mo as 1/2$^{+}$, of which configuration may be characterized by [411]1/2$^{+}$. If the multi-polarity is *E*1 or *M*1, the isomeric state in $^{109}$Mo can be interpreted as a shape isomer since the transition probability reveals a hindered value. A shape isomer is suggested in the neighboring isomer $^{109}$Nb$^{m}$ [42].

### M. $^{117}$Ru$^{m}$ and $^{119}$Ru$^{m}$

The proposed level scheme for $^{117}$Ru$^{m}$ is shown in Fig. 16. A total of five new isomeric $\gamma$-rays at 57.8-, 82.5-, 102.9-, 127.4- and 184.6-keV with a half-life of $2.487^{+0.058}_{-0.055}$ $\mu$s are observed in the case of $^{117}$Ru$^{m}$. A $\gamma\gamma$ coincidence is clearly observed between the 82.5- and 102.9-keV $\gamma$-rays. A $\gamma\gamma$ coincidence between the 57.8- and 127.4-keV $\gamma$-rays is also observed, although the statistics are low. The energy sum of these cascading $\gamma$-rays agrees with each other as well as with the energy of the 184.6-keV $\gamma$-ray transition. Inverse ordering of the cascading $\gamma$-rays is also possible.

We propose that the isomerism of $^{117}$Ru$^{m}$ is due to shape coexistence. The RUL analysis considering the branching of $\gamma$ rays allows us to restrict the multi-polarity of the 184.6-keV transition to *E*1, *M*1 or *E*2, each of which gives a hindered value for the transition probability, consistent with shape isomerism. The intensity balance analysis for the cascading 82.5- and 102.9-keV $\gamma$-rays restricts the multi-polarity of the 82.5-keV transition to *E*1 and that of the 102.9-keV transition to *E*1 or *M*1. The analysis also allows a combination of *E*2 and *M*2, but such possibility is excluded because the 102.9-keV state would have a long half-life. The 82.5-keV transition has a hindered transition probability if an *E*1 multi-polarity is assumed. These results for the 184.6- and 82.5-keV transitions support our proposed isomerism of $^{117}$Ru$^{m}$. We note that inverse ordering of the 82.5- and 102.9-keV transitions does not affect the proposed isomerism.

In the case of $^{119}$Ru$^{m}$, two new isomeric $\gamma$-rays at 90.8- and 136.3-keV with a half-life of $0.383^{+0.022}_{-0.021}$ $\mu$s are observed. We propose the level scheme shown in Fig. 16 for $^{119}$Ru$^{m}$, based on an observed $\gamma\gamma$ coincidence between the two $\gamma$-rays (See Fig. 5). The present data cannot rule out that the ordering of the 90.8- and 136.3-keV transition can be reversed. The RUL analysis allows us to restrict the multi-polarity of isomeric transition to $E1$, $M1$ or $E2$, regardless of the ordering of these $\gamma$-rays. In the case of $E1$ or $M1$, the transition probability reveals a hindered value, suggesting that $^{119}$Ru$^{m}$ is a shape isomer. However, if it is an $E2$ transition, the transition probability is not hindered. Our intensity balance analysis indicates that the hindered $E1$ or $M1$ transition is more likely, although the statistics are low. Further experimental investigations are needed to determine the isomerism of $^{119}$Ru$^{m}$ conclusively.

### N. $^{120}$Rh$^{m}$, $^{122}$Rh$^{m}$ and $^{124}$Ag$^{m}$

The following new isomeric $\gamma$-rays are observed in the delayed $\gamma$-ray energy spectra: the 59.1- and 98.1-keV $\gamma$-rays with a half-life of $0.294^{+0.016}_{-0.015}$ $\mu$s in $^{120}$Rh$^{m}$, the 63.9- and 207.1-keV $\gamma$-rays with a half-life of $0.82^{+0.13}_{-0.11}$ $\mu$s in $^{122}$Rh$^{m}$ and the 75.5- and 155.6-keV $\gamma$-rays with a half-life of $1.62^{+0.29}_{-0.24}$ $\mu$s in $^{124}$Ag$^{m}$. We propose the similar level schemes for these neighboring odd-odd isomers as shown in Fig. 17, because the $\gamma$-ray energy spectra resemble each other. In the case of $^{122}$Rh$^{m}$, a $\gamma\gamma$ coincidence is observed between the 63.9- and 207.1-keV $\gamma$-rays, although the statistics are low, supporting the proposed level scheme. In all cases the present data cannot rule out that the ordering of the transitions may be reversed. In the neutron-rich odd-odd Ag isotopes such as $^{122}$Ag [74], $\beta$-decaying long-lived isomers appear systematically at low excitation energies. The nuclide $^{124}$Ag may have such a long-lived isomer that is fed by the isomeric $\gamma$-rays observed in the present experiment.

We do not have sufficient statistics to perform an intensity balance analysis for $^{120}$Rh$^{m}$, $^{122}$Rh$^{m}$ and $^{124}$Ag$^{m}$. However, according to the RUL analysis for these isomers, $E1$, $M1$ or $E2$ are the most likely for the multi-polarity of the isomeric transition, regardless of the ordering of the $\gamma$ rays. The transition probability reveals hindered values except for in some instances the case of *E2* multi-polarity.

### O. $^{121}$Pd$^{m1,m2}$

A new isomeric $\gamma$-ray at 135.5 keV is observed in the delayed $\gamma$-ray spectrum of $^{121}$Pd.

The time spectrum shows a bump at around 500 ns, not consistent with single-component decay (see Fig. 3). Above the 135.5-keV isomeric transition, there should be unobserved isomeric transitions with a $\gamma$-ray energy below our energy threshold or with a large internal conversion coefficient, in which case the $\gamma$-ray intensity is too low to be observed in this experiment. If the ordering is reversed, the 135.5-keV transition should have a single decay component. Based on these results and also assuming two cascading-decay components, we propose the level scheme shown in Fig. 18 for $^{121}$Pd$^{m1,\,m2}$. The fitting analysis of the time spectrum allows us to determine the half-lives of the 135.5-keV transition $^{121}$Pd$^{m1}$ and the unobserved transition $^{121}$Pd$^{m2}$ as $0.460^{+0.085}_{-0.092}$ $\mu$s and $0.463^{+0.083}_{-0.094}$ $\mu$s, respectively. The population probabilities of these two isomers are determined in the fitting analysis. The RUL analysis allows us to restrict the multi-polarity of the 135.5-keV transition to *E*1, *M*1 or *E*2. In the case of *E*1 or *M*1, the transition probability reveals a hindered value, while it is not hindered if the multi-polarity is *E*2. Further experimental studies are needed to propose the level scheme conclusively.

## P. $^{125}$Ag$^{m}$

The delayed $\gamma$-ray energy and time spectra of $^{125}$Ag$^{m}$ are shown in Fig. 19 and our proposed level scheme for $^{125}$Ag$^{m}$ is shown in Fig. 20. Four isomeric $\gamma$-rays at 672, 686, 717 and 731 keV with a half-life of 473(111) ns were previously reported for $^{125}$Ag$^{m}$ and a level scheme was proposed [75]. In the present experiment seven isomeric $\gamma$-rays at 103.2, 669.8, 684.0, 714.5, 728.4, 765.1 and 786.7 keV with a half-life of $0.498^{+0.021}_{-0.020}$ $\mu$s are identified. All the previously reported $\gamma$-rays are confirmed, although the $\gamma$-ray energies systematically differ by 2-3 keV. Those at 103.2, 765.1 and 786.7 keV are new isomeric $\gamma$-rays. The energy sum of the 684.0- and 714.5-keV transitions agrees with that of 669.8- and 728.4-keV transitions, supporting the existence of the 1399-keV state. The half-life deduced for the strongest $\gamma$-ray peak at 103.2 keV is in agreement with the one derived from the summed events of the previously reported $\gamma$-rays. We place the 103.2-keV transition above the 1399-keV state as the isomeric transition, with the isomeric state located at 1502 keV. The energy sum of the 103.2-, 684.0- and 714.5-keV transitions agrees with that of the 786.7- and 714.5-keV transitions, supporting the placement of the 103.2-keV transition as well as that of the 714.5-keV transition. The 765.1-keV $\gamma$-ray cannot be assigned in the proposed level scheme.

The 1502-keV state in $^{125}$Ag$^{m}$ is likely an *E*2 isomer having spherical structure, based on the intensity balance analysis. The most consistent intensity balance is obtained between the 103.2-keV $\gamma$-ray and the two $\gamma$-rays in the cascades, if an $E2$ multi-polarity is assumed for the 103.2-keV transition. The deduced $B(E2)$ value is 1.08(12) W.u. for this isomeric transition, supporting our proposed isomerism. Here the branching of the 103.2- and 786.7-keV transitions is taken into consideration. The $J^\pi$ assignments to the energy levels in $^{125}$Ag$^m$ are tentatively made based on the theoretical predictions given in Ref. [75]. The ordering of the 669.8- and 728.4-keV transitions could be reversed.

## Q. $^{124}$Pd$^m$ and $^{126}$Ag$^m$

New isomeric $\gamma$-ray peaks are observed at 62.2 keV in $^{124}$Pd$^m$ and at 254.8 keV in $^{126}$Ag$^m$. Their half-lives are estimated to be greater than 20 $\mu$s, because the isomeric $\gamma$-ray events are equally distributed within the 20-$\mu$s range of the time spectra. The statistics are not sufficient to propose a level scheme.

## R. Other isomers

All the previously reported isomeric $\gamma$-rays in the following known isomers are identified in the present experiment: $^{43}$S$^m$ [76], $^{54}$Sc$^m$ [20, 44], $^{60}$V$^{m1,\,m2}$ [77], $^{76}$Ni$^m$ [2], $^{75}$Cu$^{m1,\,m2}$ [77], $^{78}$Zn$^m$ [3], $^{95}$Kr$^m$ [55], $^{106}$Nb$^m$ [78], $^{125}$Cd$^m$ [79], $^{127}$Cd$^m$ [80], $^{128}$Cd$^{m1,\,m2}$ [23], $^{130}$Cd$^m$ [4], $^{129}$In$^m$ [81], $^{130}$In$^m$ [82], $^{131}$In$^m$ [83], $^{130}$Sn$^m$ [84], $^{134}$Sn$^m$ [85] and $^{136}$Sb$^m$ [22, 86] as listed in Table II, and overall we find good agreement with the previous measurements, including the relative $\gamma$-ray intensities and the level schemes. The previously reported $\gamma\gamma$ coincidences in $^{78}$Zn$^m$ [3], $^{95}$Kr$^m$ [55], $^{129}$In$^m$ [81] and $^{134}$Sn$^m$ [85] are confirmed. A $\gamma\gamma$ coincidence is observed between the 53.7- and 173.0-keV $\gamma$-rays in $^{136}$Sb$^m$, supporting the level scheme proposed in Ref. [22, 86]. The relative $\gamma$-ray intensities are deduced for $^{60}$V$^{m1,\,m2}$ [85], $^{75}$Cu$^{m1,\,m2}$ [77] and $^{106}$Nb$^m$ [78] for the first time. In the case of $^{60}$V$^{m1,\,m2}$, the observed intensity of the upper isomeric transition $^{60}$V$^{m2}$ is approximately two times higher than that of the lower transition $^{60}$V$^{m1}$ (see Table III). This implies the possibility of reversing the previously reported ordering of these two transitions [77].

Overall agreement with the previous measurements is also achieved in the case of the following known isomers: $^{50}$K$^m$ [77], $^{56}$Sc$^m$ [44], $^{64}$Mn$^m$ [20, 77, 87], $^{67}$Fe$^m$ [88], $^{96}$Rb$^m$ [56], $^{97}$Sr$^m$ [57, 58], $^{100}$Sr$^m$ [10], $^{101}$Y$^m$ [27], $^{128}$Cd$^{m3}$ [23], $^{131}$Sn$^m$ [89] and $^{132}$Sn$^m$ [90], although some of the previously reported $\gamma$-rays are not observed as seen in Table II. This disagreement on observed $\gamma$-ray peaks can be attributed to our limited statistics and energy threshold. We find disagreement with the previously reported half-life of $^{101}$Y$^m$ [27]: The present measurement is approximately five times shorter than the previously reported value.

## IV. SUMMARY AND CONCLUSIONS

The present experiment includes the measurement of new and known isomers with $N$ ~ 58 to 61 and $Z$ ~ 34 to 39. It provides us with the opportunity to observe and elucidate the evolution of shape coexistence in the corresponding region of neutron-rich exotic nuclei. Previously it was shown that prolate-deformed states coexist with spherical states in this region [6]. In the present work we propose that the isomers $^{95}Br^{m}$, $^{97}Rb^{m}$ and $^{98}Rb^{m}$ are shape isomers generated due to such shape coexistence. Furthermore the energy difference between the prolate-deformed and spherical states systematically gets smaller as the proton number decreases along the $N$ = 60 line and their ordering could even be reversed. The isomeric state at 77.1 keV in $^{97}Rb^{m}$ is particularly interesting, because the 77.1-keV state and the ground state are located very close to each other, implying that the $^{97}Rb$ is likely a transitional nucleus. The ordering of the prolate-deformed and spherical states is reversed in our proposed level scheme of $^{95}Br^{m}$. Further investigations including theoretical studies, particularly for the Se and Br isotopes, are needed to fully understand the evolution of shape coexistence and the nuclear isomerism in this region.

In the region of $N$ ~ 65 to 70 and $Z$ ~ 40 to 43, a variety of nuclear shapes and their coexistence are involved in the appearance of the isomers observed in the present experiment. The nuclear isomerism of $^{108}Zr^{m}$ was previously discussed in terms of a tetrahedral shape [43]. In the present work we propose the existence of a $K$ isomer for $^{108}Zr^{m}$ instead, based on our expended level scheme and the recent theoretical prediction [71]. A low-lying oblate shape isomer was proposed in Ref. [42] in the case of $^{109}Nb^{m}$, where the ground state is known to be of a prolate shape. Triaxial-deformed shape isomers were proposed above oblate-deformed states in the cases of $^{112}Tc^{m}$ and $^{113}Tc^{m}$ [41]. In the present work we propose two possibilities for the new isomer $^{108}Nb^{m}$: an $E2$ isomer caused by a small transition-energy between deformed states in different bands or an oblate-deformed shape isomer located above a prolate-deformed state. We also propose that the new isomer $^{109}Mo^{m}$ is an $E2$ isomer resulting from the transition between deformed band heads, in analogy to the isomerism of neighboring $^{107}Mo^{m}$ [11]. These results indicate a rich variety of phenomena in this region and that more comprehensive information is needed to fully elucidate the nuclear structure.

The isomers in the region of $N$ ~ 75 and Z ~ 44 to 47 were not explored in any detail prior to the present experiment. The present observation of the new isomers in $^{117}$Ru$^{m}$, $^{119}$Ru$^{m}$, $^{120}$Rh$^{m}$, $^{122}$Ru$^{m}$, $^{121}$Pd$^{m}$, $^{124}$Pd$^{m}$, $^{124}$Ag$^{m}$ and $^{126}$Ag$^{m}$ provide us with clues to understanding isomerism and the corresponding nuclear structure in this region. We propose $^{117}$Ru$^{m}$ to be a shape isomer, based on the hindrance of its transition probability. The possible isomerism for $^{119}$Ru$^{m}$, $^{120}$Rh$^{m}$, $^{122}$Ru$^{m}$, $^{121}$Pd$^{m}$ and $^{124}$Ag$^{m}$ is also discussed. The $\gamma$-ray energy spectra of the observed isomers are similar to each other, implying the presence of a common mechanism. The occurrence of nuclear deformation in the region of Z ~ 39 to 45 and $N$ ~75 was pointed out in Ref. [15] to be of astrophysical interest. The mass model based on the Extended Thomas-Fermi plus Strutinsky Integral (ETFSI) method indicates the occurrence of deformation in the relevant region [91]: the systematics of two-neutron separation energies exhibit humps around $N$ ~ 75 in the isotopic chains of $Z$ ~ 39 to 45 (see Fig. 3(c) in Ref. [91]). This behavior is very similar to what was observed around $N$ ~ 60 in the isotopic chain of $Z$ ~ 38 to 42 [62], where shape coexistence systematically appears. This could give rise to the shape isomers observed in the present work. We speculate that the isomers in the region of $N$ ~ 75 could be generated in a similar way. The relevant neutron-rich region could be a new deformed region exhibiting shape coexistence. Further experimental and theoretical studies are needed to understand the nuclear structure and isomerism in this neutron-rich region.

Finally, isomers in a lighter-mass region with $N$ ~ 28 to 50 are also observed and investigated in the present work, providing us with an opportunity to study shell evolution in the relevant neutron-rich exotic nuclei. For instance, the existence of the isomer $^{59}$Ti$^{m}$ suggests that the $N$ = 34 sub-shell gap between the $\nu p_{1/2}$ and $\nu f_{5/2}$ orbitals gets smaller as the neutron number increases. Such shell evolution can be attributed to the attractive monopole interaction between the $\pi f_{7/2}$ and $\nu f_{5/2}$ orbitals [92].

In summary, we have conducted a search for new isomers using in-flight fission of a $^{238}$U beam at 345 MeV/nucleon, and a number of neutron-rich microsecond isomers were identified over a wide range of atomic numbers, including the discovery of eighteen new isomers. A wealth of spectroscopic information was obtained, allowing us to present proposed level schemes for seventeen neutron-rich exotic nuclei far from stability. In the present experiment we were able to explore nuclear isomerism over a wide range of neutron-rich exotic nuclei, allowing the investigation of the evolution of nuclear structure and shape coexistence. The experiment was made possible by the use of in-flight fission as well as the features of BigRIPS separator which were fully exploited. These results demonstrate the potential of RIBF, which will allow isomer spectroscopy over a wide range of neutron-rich exotic nuclei far beyond the limits of conventional means.

**ACKNOWLEDGEMENTS**

The present experiment was carried out under Program Number NP0702-RIBF20 at the RI Beam Factory operated by RIKEN Nishina Center, RIKEN and CNS, University of Tokyo. The authors are grateful to the RIBF accelerator crew for providing the uranium beam. They also would like to thank Dr. Y. Yano, RIKEN Nishina Center, for his support and encouragement. The authors SM, JN were supported by the U.S. Department of Energy, Office of Nuclear Physics, under Contract No. DE-AC02-06CH11357. The authors AN, OT, DB, BS were supported by the National Science Foundation under Grant No. PHY-0606007 and by the U.S. Department of Energy, Office of Nuclear Physics, under Grant No. DE-FG02-03ER41265. The Author MF was supported by the National Science Foundation under Grants No. PHY-0855013 and PHY-0735989. The author TK is grateful to Prof. J. Kasagi and Dr. J. Stasko for their careful reading the manuscript.

TABLE I. New isomers observed in the present work along with the corresponding BigRIPS settings, numbers of implanted fragments, half-lives ($T_{1/2}$), $\gamma$-ray energies ($E_\gamma$) and $\gamma$-ray relative intensities ($I_\gamma$).

| Isomers | BigRIPS setting [a] | Number of fragments | $T_{1/2}$ [$\mu$s] | $E_\gamma$ [keV] | $I_\gamma$ [%] [b] |
|---|---|---|---|---|---|
| $^{59}Ti^m$ | G1 | $1.1 \times 10^4$ | $0.587^{+0.057}_{-0.051}$ | 109.0 | - |
| $^{90}As^m$ | G1 | $1.1 \times 10^4$ | $0.20^{+0.12}_{-0.09}$ | 124.5 | - |
| $^{92}Se^m$ | G1 | $6.4 \times 10^4$ | $10.3^{+5.5}_{-2.8}$ [c] | 503.4 | 100(20) |
| | | | | 538.8 | 89(18) |
| | | | | 897.8 | 76(18) |
| $^{93}Se^m$ | G1 | $3.7 \times 10^3$ | $0.39^{+0.12}_{-0.08}$ | 208.3 | 85(26) |
| | | | | 469.9 | 100(43) |
| $^{94}Br^m$ | G1,G2 | $2.4 \times 10^5$ | 0.530(15) | 88.0 | 2.6(20) |
| | | | | 92.1 | 44(3) |
| | | | | 115.5 | 6.2(13) |
| | | | | 165.7 | 12(2) |
| | | | | 179.0 | 15(2) |
| | | | | 202.6 | 100(4) |
| $^{95}Br^m$ | G1,G2 | $5.0 \times 10^4$ | $6.7^{+1.1}_{-0.9}$ [c] | 537.9 | - |
| $^{96}Br^m$ | G1,G2 | $4.1 \times 10^3$ | $2.7^{+1.1}_{-0.7}$ [c] | 311.5 | - |
| $^{97}Rb^m$ | G1,G2 | $1.4 \times 10^6$ | $6.33^{+0.37}_{-0.34}$ [c,d] | 77.1 | - |
| $^{108}Nb^m$ | G2 | $2.7 \times 10^6$ | 0.109(2) [e] | 64.3 | 1.6(4) |
| | | | | 77.6 | 100(2) |
| | | | | 89.0 | 56(1) |
| | | | | 102.2 | 19(1) |
| $^{109}Mo^m$ | G2 | $8.3 \times 10^3$ | $0.194^{+0.076}_{-0.049}$ | 69.7 | - |
| $^{117}Ru^m$ | G2 | $3.1 \times 10^5$ | $2.487^{+0.058}_{-0.055}$ [f] | 57.8 | 3.9(4) |
| | | | | 82.5 | 28(1) |
| | | | | 102.9 | 27(1) |
| | | | | 127.4 | 5.1(5) |
| | | | | 184.6 | 100(2) |
| $^{119}Ru^m$ | G2 | $1.6 \times 10^4$ | $0.383^{+0.022}_{-0.021}$ | 90.8 | 69(6) |
| | | | | 136.3 | 100(7) |
| $^{120}Rh^m$ | G2 | $1.7 \times 10^5$ | $0.294^{+0.016}_{-0.015}$ | 59.1 | 8.2(15) |
| | | | | 98.1 | 100(5) |
| $^{122}Rh^{m}$ | G2 | $8.6 \times 10^{3}$ | $0.82^{+0.13}_{-0.11}$ | 63.9 | 35(9) |
| | | | | 207.1 | 100(17) |
| $^{121}Pd^{m1}$ [g] | G2 | $2.8 \times 10^{4}$ | $0.460^{+0.085}_{-0.092}$ | 135.5 | - |
| $^{121}Pd^{m2}$ [h] | | | $0.463^{+0.083}_{-0.094}$ [i] | - | - |
| $^{124}Pd^{m}$ | G2 | $3.8 \times 10^{4}$ | > 20 [j] | 62.2(16) [k] | - |
| $^{124}Ag^{m}$ | G2,G3 | $5.7 \times 10^{4}$ | $1.62^{+0.29}_{-0.24}$ | 75.5 | 32(7) |
| | | | | 155.6 | 100(12) |
| $^{126}Ag^{m}$ | G2,G3 | $1.3 \times 10^{5}$ | > 20 [j] | 254.8 | - |

[a] G1, G2 and G3 indicate the BigRIPS settings that were tuned for the Z~30, Z~40 and Z~50 regions, respectively (see text). Experimental conditions are given in Ref. [32].

[b] The relative intensities were derived from photo-peak counts which were corrected with the $\gamma$-ray detection efficiency. They are normalized to the intensity of the strongest one among the observed peaks. The errors shown are statistical only. We estimate that the $\gamma$-ray detection efficiency has systematic errors of 15%.

[c] In this case the constant-background component in the time spectrum was separately deduced by fitting the neighboring events on the high-energy side of the relevant $\gamma$-ray peak, because the long half-life relative to the TDC time range did not allow the fitting analysis with a decay component plus a constant background. The error shown also includes the fitting error of the constant-background component.

[d] Only the G2 data were used to deduce the half-life, because its signal-to-noise ratio was much better than that of the G1 data. This is due to the smaller $\gamma$-ray attenuation in the stopper (see text).

[e] The half-life was deduced by using the most intense $\gamma$-ray at 77.6 keV.

[f] The half-life was deduced by using the most intense $\gamma$-ray at 184.6 keV.

[g] $m1$ represents the isomer located at the first lowest excitation energy.

[h] $m2$ represents the isomer located at the second lowest excitation energy.

[i] The half-life of the unobserved $^{121}Pd^{m2}$ was deduced by assuming two cascading-decay components and a constant background for the time spectrum of the observed 135.5-keV $\gamma$-ray (see text).

[j] The half-life was estimated as given here, because the isomeric $\gamma$-ray events were equally distributed in the 20-$\mu$s range of the time spectrum.

[k] The statistical error is given in the parentheses.

TABLE II. $\gamma$-ray energies and half-lives of previously known isomers observed in the present work. The BigRIPS settings and numbers of implanted fragments are also displayed. The previously-reported $\gamma$-ray energies and half-lives are presented for comparison along with the references. New experimental information obtained in the present work is indicated by boldface with an underline.

| Isomers | BigRIPS setting [a] | Number of fragments | $E_\gamma$ [keV] | $T_{1/2}$ [$\mu$s] | Literature | | |
|---|---|---|---|---|---|---|---|
| | | | | | $E_\gamma$ [keV] | $T_{1/2}$ [$\mu$s] | Ref. |
| $^{43}S^{m}$ | G1 | $4.6 \times 10^3$ | 320.9 | $0.20^{+0.14}_{-0.07}$ | 319 | 0.478(48) | [76] |
| $^{50}K^{m}$ | G1 | $2.0 \times 10^4$ | 128.5;172.4 | $0.138^{+0.050}_{-0.041}$ | 44;70;101;127.4;171.4 | 0.125(40) | [77] |
| $^{54}Sc^{m}$ | G1 | $2.8 \times 10^4$ | 110.7 | $2.78^{+0.31}_{-0.26}$ | 110.5 | 2.77(2) | [20, 44] |
| $^{56}Sc^{m}$ | G1 | $8.0 \times 10^3$ | 140.6;188.2 | $0.35^{+0.26}_{-0.12}$ | 47.7;140.5;187.8;587.2;727.1 | 0.290(30) | [44] |
| $^{60}V^{m1}$ [b] | G1 | $1.1 \times 10^5$ | 104.0 | - [d] | 103.2 | 0.013(3) | [77] |
| $^{60}V^{m2}$ [c] | G1 | | 99.7;104.0 | $0.229^{+0.025}_{-0.023}$ | 98.9;103.2 | 0.320(90) | [77] |
| $^{64}Mn^{m}$ | G1 | $1.6 \times 10^5$ | 135.3 | - | 39.8;134.3 | >100 | [20, 77] |
| | | | | | 40.2;135.3 | 400(40) | [87] |
| $^{67}Fe^{m}$ | G1 | $2.2 \times 10^5$ | 366.8 | - | 366.4;387.7 | 64(14) | [88] |
| $^{76}Ni^{m}$ | G1 | $1.1 \times 10^4$ | 142.7;355.5;929.9;990.6 | $0.409^{+0.058}_{-0.050}$ | 144;354;930;992 | $0.590^{+0.180}_{-0.110}$ | [2] |
| $^{75}Cu^{m1}$ [b] | G1 | $5.4 \times 10^5$ | 62.5 | - [e] | 61.8 | 0.370(40) | [77] |
| $^{75}Cu^{m2}$ [c] | G1 | | 66.8;62.5 | $0.134^{+0.025}_{-0.020}$ | 66.5;61.8 | 0.170(15) | [77] |
| $^{78}Zn^{m}$ | G1 | $1.3 \times 10^6$ | 145.7;730.0;890.5;909.1 | $0.320^{+0.009}_{-0.008}$ | 144.7;729.6;889.9;908.3 | 0.319(9) | [3] |
| $^{82}Ga^{m}$ | G1 | $1.1 \times 10^6$ | 141.0 | $\mathbf{\underline{0.098}^{+0.010}_{-0.009}}$ | 141 | < 0.5 | [27] |
| $^{92}Br^{m}$ | G1 | $1.7 \times 10^6$ | 98.6;106.3;139.7;154.9;**197.6**;239.5;259.5;301.9;778.8;**898.3**;1039.8 | $\mathbf{\underline{0.089}^{+0.007}_{-0.008}}$ [f]; $\mathbf{\underline{0.084}^{+0.010}_{-0.009}}$ [f] | 99;106;139;155;169;239;259;295;301;780;1039 | < 0.5 | [27] |
| $^{95}Kr^{m}$ | G1.G2 | $1.7 \times 10^6$ | 82.6;114.3 | 1.582(22) | 81.7;113.8 | 1.4(2) | [55] |
| $^{96}Rb^{m}$ | G2 | $6.4 \times 10^4$ | 59.9;90.3;93.3;117.2;123.2 [h];149.7;177.2;185.1;210. | $1.72^{+0.16}_{-0.14}$ [g] | 38.0;59.3;89.5;92.8;116.8;122.0;123.5;126.0;148.8;16 | 2.00(10) | [56] |
| | | | 2;240.5;300.2 [h];368.5 [h];461.4 | | 6.1;177.6;185.4;209.9;240.3;276.3;300.0;301.0;329.0;366.8;369.2;402.4;405.5;461.6;495.2;554.5 | | |
| $^{98}Rb^{m}$ | G1,G2 | $1.7 \times 10^{6}$ | 123.7;178.4 | 0.358(7) | 116;124;178 | $0.700^{+0.060}_{-0.050}$ | [27] |
| $^{97}Sr^{m1}$ [b] | G2 | $3.2 \times 10^{3}$ | 141.3;167.6 | $0.52^{+0.16}_{-0.12}$ | 140.8;167.0 | 0.170(10) | [57] |
| $^{97}Sr^{m2}$ [c] | G2 | | - | - | 140.8;167.0;522.0 | 0.265(27) | [57] |
| | | | - | - | 141.0;167.1;522.1 | 0.526(13) | [58] |
| $^{100}Sr^{m}$ | G2 | $1.6 \times 10^{6}$ | 129.6;288.1;1201.8 | 0.122(9) | 129.3;204.0;288.2;997.5;1201.0;1285.5;1370.7 | 0.085(7) | [10] |
| $^{101}Y^{m}$ | G2 | $1.7 \times 10^{5}$ | 128.0; 203.5 | $0.187^{+0.049}_{-0.038}$ | 129;164;204;230;480 | $0.860^{+0.090}_{-0.080}$ | [27] |
| $^{108}Zr^{m}$ | G2 | $1.5 \times 10^{5}$ | 174.4;277.9;347.2;**364.2**;**426.3**;**432.3**;478.3;**485.3**;603.9;**642.3**;**773.6**;**795.9**;**828.3** | $0.536^{+0.026}_{-0.025}$ | 173.7;278.6;347.9;477.5;605.6 | 0.620(150) | [43] |
| $^{106}Nb^{m}$ | G2 | $1.3 \times 10^{4}$ | 63.5;94.7;108.1;147.5;202.1;204.9 | $0.66^{+0.11}_{-0.10}$ | 63.5;94.5;107.3;147.4;201.8;204.6 | 0.84(4) | [78] |
| $^{109}Nb^{m}$ | G2 | $1.6 \times 10^{6}$ | 117.0;196.2;312.2 | $0.114^{+0.008}_{-0.007}$ [i] | 117;196.3;313.1 | 0.150(30) | [42] |
| $^{112}Tc^{m}$ | G2 | $7.2 \times 10^{3}$ | 93.1;259.2 | $0.218^{+0.060}_{-0.043}$ | 92;258 | 0.150(17) | [27, 41] |
| $^{113}Tc^{m}$ | G2 | $6.7 \times 10^{5}$ | 114.4 | $0.526^{+0.016}_{-0.015}$ | 114 | 0.5(1) | [41] |
| $^{125}Ag^{m}$ | G2,G3 | $2.4 \times 10^{5}$ | **103.2**;669.8;684.0;714.5;728.4;**765.1**;**786.7** | $0.498^{+0.021}_{-0.020}$ | 672.1;686.3;716.6;731.2 | 0.473(111) | [75] |
| $^{125}Cd^{m}$ | G3 | $2.6 \times 10^{3}$ | 719.4;742.2 | - | 719.1;742.7 | 19(3) | [79] |
| $^{127}Cd^{m}$ | G2,G3 | $2.5 \times 10^{5}$ | 110.6;710.1;738.5;821.1;848.8 | $11.0^{+9.2}_{-3.5}$ [k,l] | 110.4;711.2;738.8;821.2;848.9 | 17.5(3) | [80] |
| $^{128}Cd^{m1}$ [b] | G2,G3 | $6.5 \times 10^{5}$ | 439.2;644.9;783.9;1223.7 | - [e] | 440.0;645.8;784.6;1224.0 | 0.270(7) | [23] |
| $^{128}Cd^{m2}$ [c] | | | 237.5;439.2;644. | - [d] | 237.9;440.0;645.8 | 0.012(2) | [23] |
| | | | 9;783.9;1223.7 | | ;784.6;1224.0 | | |
| $^{128}$Cd$^{m3}$ [j] | | | 68.8;237.5;439.2;536.8;644.9;783.9;1223.7 | $3.76^{+0.44}_{-0.37}$ | 68.7;237.9;440.0;450.4;537.6;645.8;765.0;784.6;1224.0 | 3.56(6) | [23] |
| $^{130}$Cd$^{m}$ | G2,G3 | $9.1 \times 10^4$ | 128.0;138.0;538.2;1325.4 | $0.248^{+0.021}_{-0.019}$ | 128;138;539;1325 | 0.220(30) | [4] |
| $^{129}$In$^{m}$ | G3 | $3.0 \times 10^5$ | 333.8;359.0;995.3;1354.6 | $11.3^{+2.2}_{-1.6}$ [k] | 333.8;359.0;995.2;1354.1 | 8.7(7) | [81] |
| $^{130}$In$^{m}$ | G2,G3 | $1.8 \times 10^6$ | 388.5 | $5.25^{+0.40}_{-0.35}$ [k] | 388.8 | 3.1(3) | [82] |
| $^{131}$In$^{m}$ | G3 | $2.0 \times 10^6$ | 3783.6 | $0.685^{+0.042}_{-0.039}$ | 3782 | 0.630(60) | [83] |
| $^{130}$Sn$^{m}$ | G3 | $2.5 \times 10^4$ | 96.7;391.6 | $1.14^{+0.27}_{-0.21}$ | 96.54;391.39 | 1.61(15) | [84] |
| $^{131}$Sn$^{m}$ | G3 | $1.0 \times 10^6$ | 158.8;173.4;344.0;4273.8 | $0.309^{+0.024}_{-0.023}$ | 158.57;173.28;344.40;4102.3;4273.2;4446.0 | 0.300(20) | [89] |
| $^{132}$Sn$^{m}$ | G3 | $2.0 \times 10^7$ | 132.7;299.4;374.8;4041.2;4416.7 | 2.088(17) [m] | 64.4;132.5;299.6;310.7;375.1;4041.1;4351.9;4416.2 | 2.03(4) | [90] |
| $^{134}$Sn$^{m}$ | G3 | $3.8 \times 10^6$ | 174.1;347.1;725.1 | $0.086^{+0.008}_{-0.007}$ | 174.0;347.8;725.6 | 0.080(15) | [85] |
| $^{136}$Sb$^{m}$ | G3 | $4.0 \times 10^6$ | 53.9;173.1 | 0.570(5) | 51.4 [n];53.4;173.0 | 0.565(50) | [22, 86] |

[a] G1, G2 and G3 indicate the BigRIPS settings that were tuned for the Z~30, Z~40 and Z~50 regions, respectively (see text). Each experimental condition is given in Ref. [32].

[b] *m*1 represents the isomer located at the first lowest excitation energy.

[c] *m*2 represents the isomer located at the second lowest excitation energy.

[d] The half-life was too short to be deduced in the present work.

[e] The half-life could not be deduced in the present work because two decay-component fitting was difficult due to the low statistics.

[f] The half-life was deduced to be 0.089 $\mu$s for the 662-keV isomer using the $\gamma$-rays at 106.3, 154.9 and 301.9 keV and 0.084 $\mu$s for the 1138-keV isomer using those at 778.8, 898.3 and 1039.8 keV. See text.

[g] This value was deduced using the $\gamma$-rays at 93.3, 117.2, 123.2, 210.2, 240.5, 300.2, 368.5 and 461.4 keV.

[h] This should be a doublet peak. See the previous report.

[i] The 196.2-keV $\gamma$-ray was not used in deducing the half-life due to the mixture of the background $\gamma$-ray at 197 keV (see text).

[j] $m3$ represents the isomer located at the third lowest excitation energy.

[k] In this case the constant-background component in the time spectrum was separately deduced by fitting the neighboring events on the high-energy side of the relevant $\gamma$-ray peak, because the long half-life relative to the TDC time range did not allow the fitting analysis with a decay component plus a constant background. The error shown also includes the fitting error of the constant-background component.

[l] The half-life was deduced using the $\gamma$-rays at 738.5 and 821.1 keV.

[m] This value was deduced using the most intense $\gamma$-ray at 132.7 keV.

[n] Only conversion electrons were observed for the 51.4-keV transition in $^{136}Sb^{m}$.

TABLE III. Deduced $\gamma$-ray relative intensities for previously reported isomers. New $\gamma$-ray peaks observed in the present work are indicated by boldface with an underline.

| Isomers | $E_\gamma$[keV] | $I_\gamma$ [%] [a] | Isomers | $E_\gamma$[keV] | $I_\gamma$ [%] [a] |
|---|---|---|---|---|---|
| $^{60}V^{m1}$ [b] | 104.0 | 60(8) [d] | ($^{109}Nb^{m}$) | 196.2 | 100(13) [g] |
| $^{60}V^{m2}$ [c] | 99.7 | 100(11) [d] | | 312.2 | 76(10) |
| $^{75}Cu^{m1}$ [b] | 62.5 | 100(14) [e] | $^{108}Zr^{m}$ | 174.4 | 100(8) |
| $^{75}Cu^{m2}$ [c] | 66.8 | 55(9) [e] | | 277.9 | 37(6) |
| $^{92}Br^{m}$ [f] | 98.6 | 100(5) | | 347.2 | 40(6) |
| | 106.3 | 59(5) | | **<u>364.2</u>** | 25(6) |
| | 139.7 | 10(6) | | **<u>426.3</u>** | 47(7) |
| | 154.9 | 66(4) | | **<u>432.3</u>** | 22(6) |
| | **<u>197.6</u>** | 2.1(36) [g] | | 478.3 | 33(7) |
| | 239.5 | 9.6(21) | | **<u>485.3</u>** | 15(5) |
| | 259.5 | 60(4) | | 603.9 | 50(9) |
| | 301.9 | 12(3) | | **<u>642.3</u>** | 22(6) |
| | 778.8 | 11(3) | | **<u>773.6</u>** | 21(7) |
| | **<u>898.3</u>** | 6.1(19) | | **<u>795.9</u>** | 14(5) |
| | 1039.8 | 20(3) | | **<u>828.3</u>** | 22(8) |
| $^{98}Rb^{m}$ | 123.7 | 100(2) | $^{125}Ag^{m}$ | **<u>103.2</u>** | 70(5) |
| | 178.4 | 45(2) | | 669.8 | 77(9) |
| $^{106}Nb^{m}$ | 63.5 | 40(13) | | 684.0 | 76(9) |
| | 94.7 | 67(19) | | 714.5 | 100(11) |
| | 108.1 | 40(15) | | 728.4 | 76(10) |
| | 147.5 | 16(12) | | **<u>765.1</u>** | 14(4) |
| | 202.1 | 66(19) | | **<u>786.7</u>** | 9.0(40) |
| | 204.9 | 100(24) | $^{136}Sb^{m}$ | 53.9 | 1.7(1) |
| $^{109}Nb^{m}$ [h] | 117.0 | 96(9) | | 173.1 | 100(1) |

[a] The relative intensities were derived from photo-peak counts which were corrected with the $\gamma$-ray detection efficiency. They are normalized to the intensity of the strongest one among the observed peaks. The errors shown are statistical only. We estimate that the $\gamma$-ray detection efficiency has systematic errors of 15%.

[b] *m*1 represents the isomer located at the first lowest excitation energy.

[c] $m2$ represents the isomer located at the second lowest excitation energy.
[d] The time window was set from 0.2 to 1.0 $\mu$s.
[e] The time window was set from 0.2 to 2.1 $\mu$s.
[f] The time window was set from 0.15 to 0.5 $\mu$s.
[g] The contribution of the 197.1-keV $\gamma$-ray originating from the $^{19}$F(n,n'$\gamma$) reaction was subtracted based on the 197.1-keV peak observed in the neighboring isotopes: $^{90}$Se for $^{92}$Br$^{m}$ and $^{106}$Zr for $^{109}$Nb$^{m}$, applying the same time window.
[h] The time window was set from 0.2 to 0.6 $\mu$s.

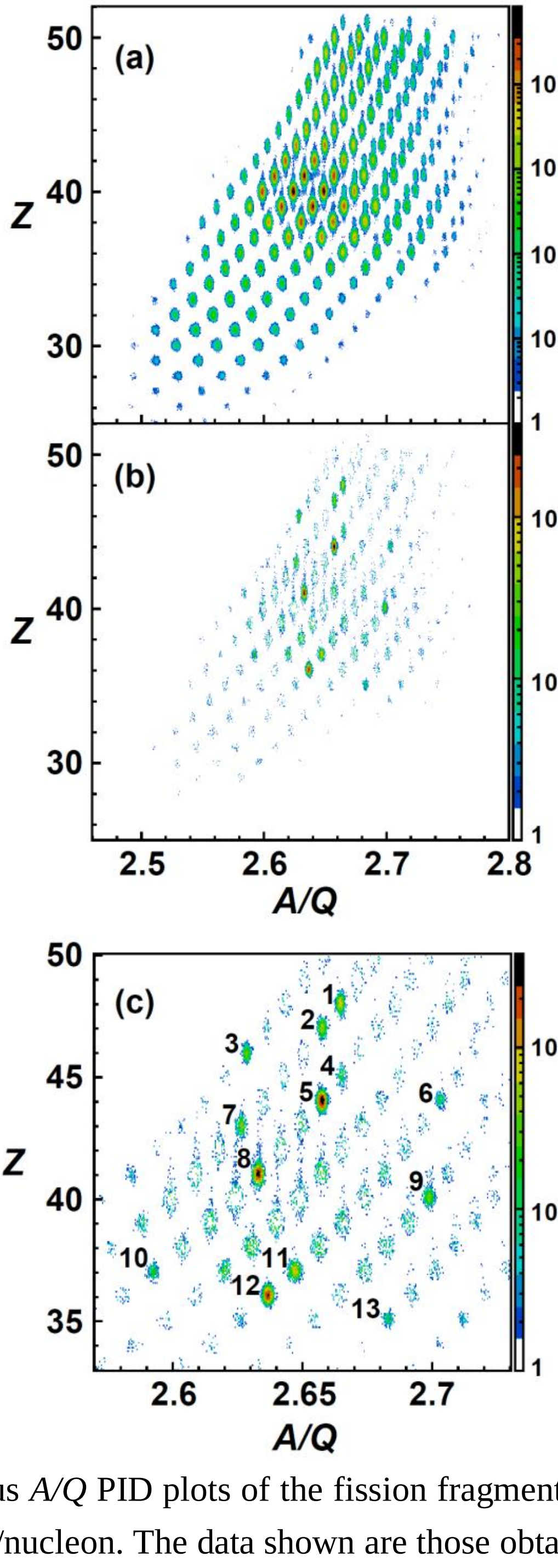

FIG. 1. (Color) $Z$ versus $A/Q$ PID plots of the fission fragments produced in the $^{238}U$ + Be reaction at 345MeV/nucleon. The data shown are those obtained with the G2 setting. The figures (a) and (b) correspond to the PID plots without and with the delayed $\gamma$ coincidence, respectively. The figure (c) shows an enlarged plot of (b), in which the strongly observed isomers are labeled as 1:$^{128}Cd^m$, 2:$^{125}Ag^m$, 3:$^{121}Pd^m$, 4:$^{120}Rh^m$, 5:$^{117}Ru^m$, 6:$^{119}Ru^m$, 7:$^{113}Tc^m$, 8:$^{108}Nb^m$, 9:$^{108}Zr^m$, 10:$^{96}Rb^m$, 11:$^{98}Rb^m$, 12:$^{95}Kr^m$ and 13:$^{94}Br^m$. The time gate used in the figures (b) and (c) is from 0.2 to 5 $\mu$s following ion implantation. See text.

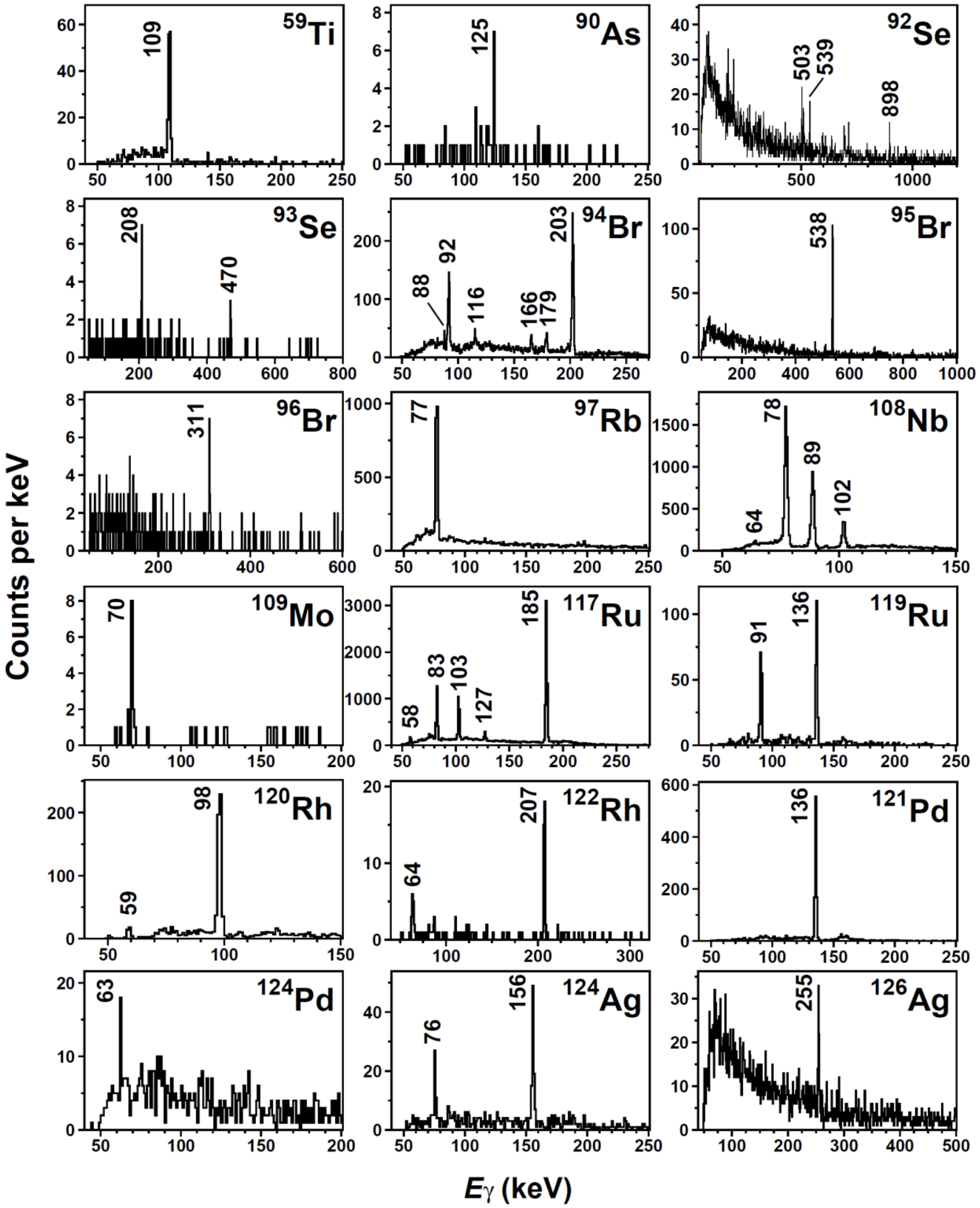

FIG. 2. Delayed $\gamma$-ray energy spectra of new isomers identified in the present work. Peaks are labeled by energies in keV. Unlabeled peaks such as those seen in $^{92}$Se are the background $\gamma$-rays described in text or unidentified $\gamma$-rays. Note that the spectra of $^{94}$Br and $^{108}$Nb are given in 0.5 keV/channel.

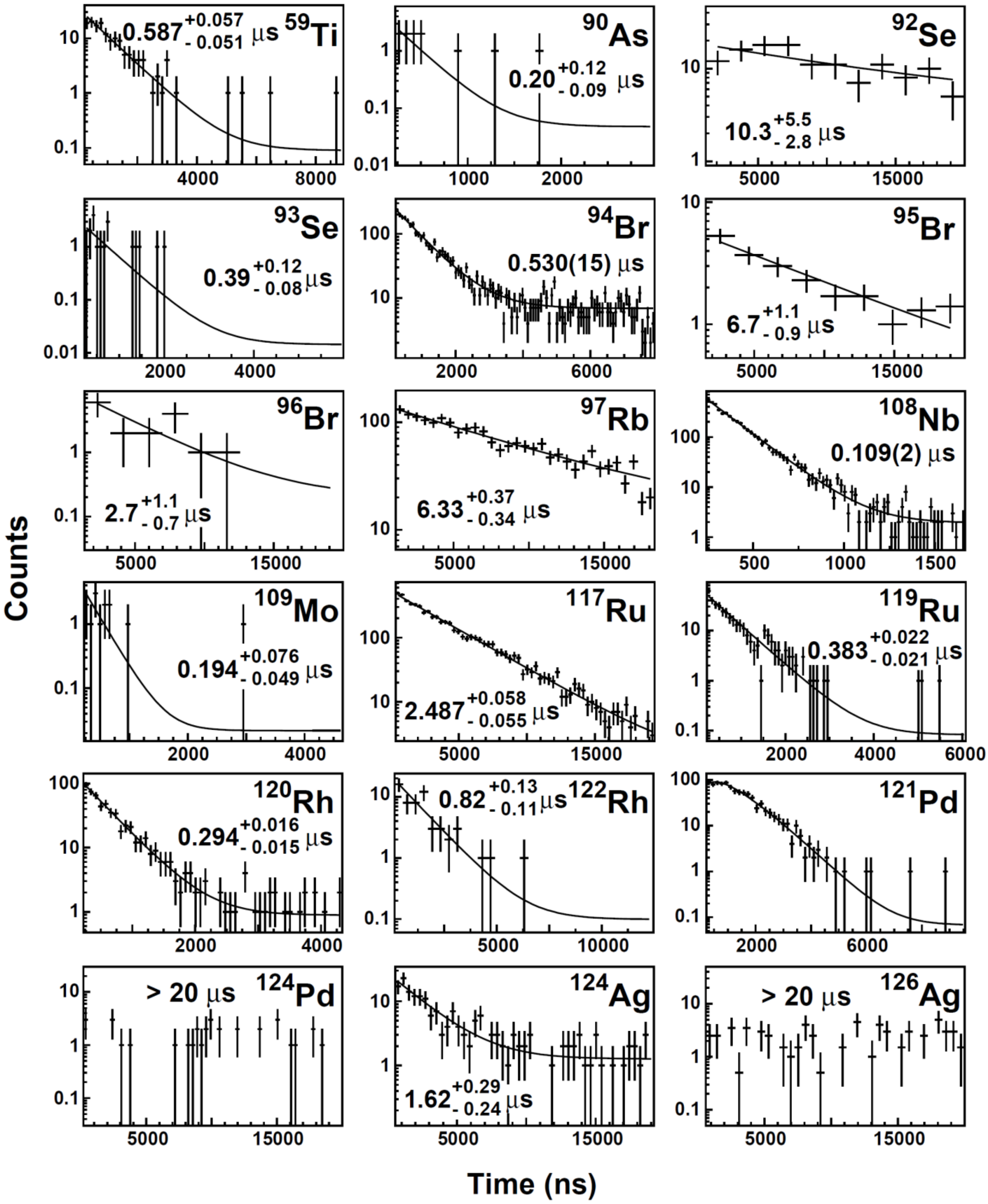

FIG. 3. Time spectra of new isomers identified in the present work. The solid line shows the fitted decay curve using an exponential decay component plus a constant background. The deduced half-lives are given in the figures except for $^{121}$Pd$^{m}$. Note that the time spectrum of $^{121}$Pd$^{m}$ is fit using two cascading-decay components plus a constant background. See text for the half-lives of $^{121}$Pd$^{m}$.

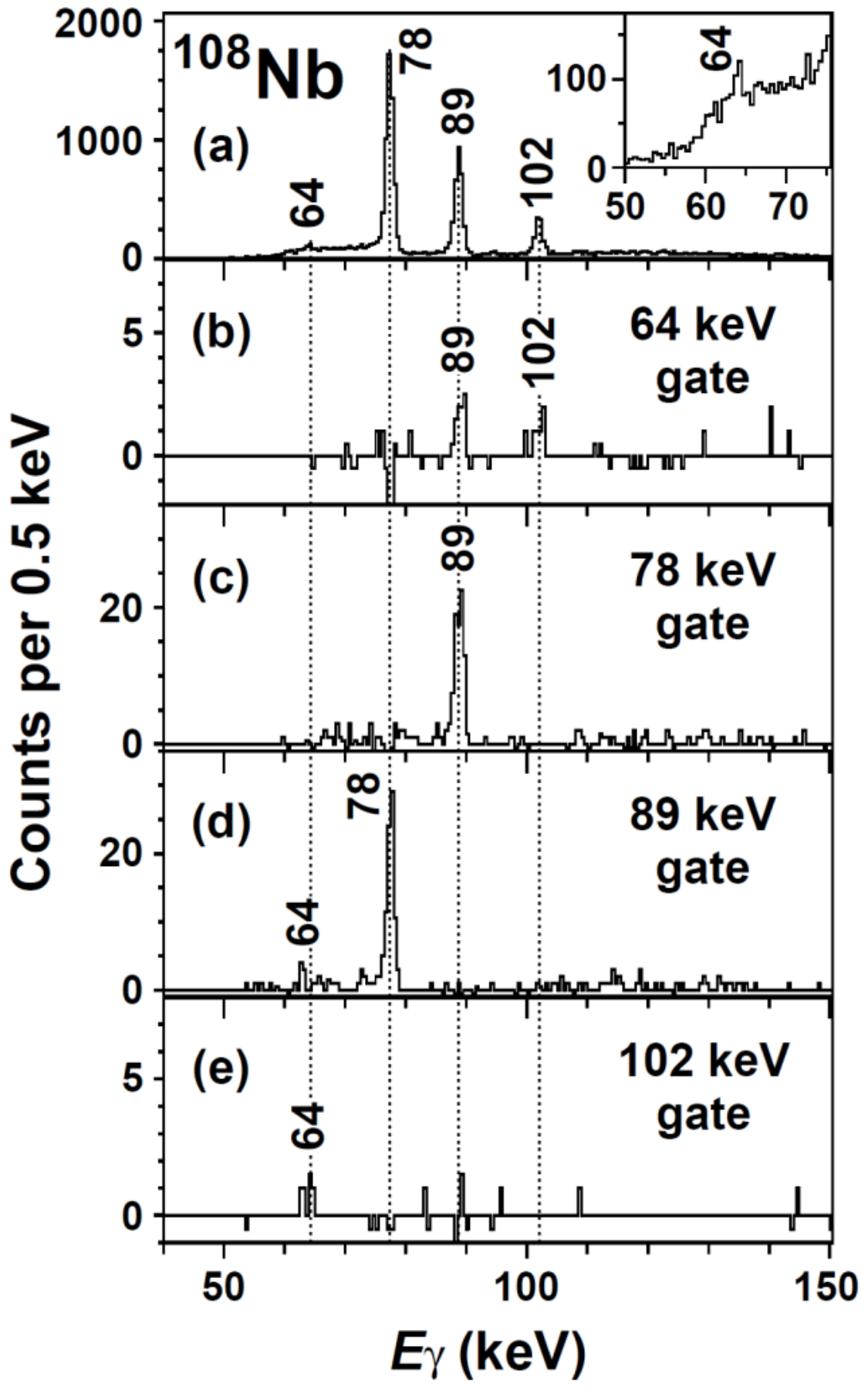

FIG. 4. Delayed $\gamma$-ray singles spectrum and $\gamma\gamma$ coincidence spectra of $^{108}Nb^m$. The $\gamma$-ray spectra observed in coincidence with (b) the 64.3-, (c) 77.6-, (d) 89.0- and (e) 102.2-keV $\gamma$-rays are shown along with (a) the $\gamma$-ray singles spectrum. A partially-enlarged $\gamma$-ray singles spectrum is shown in the inset of (a). Peaks are labeled by energies in keV.

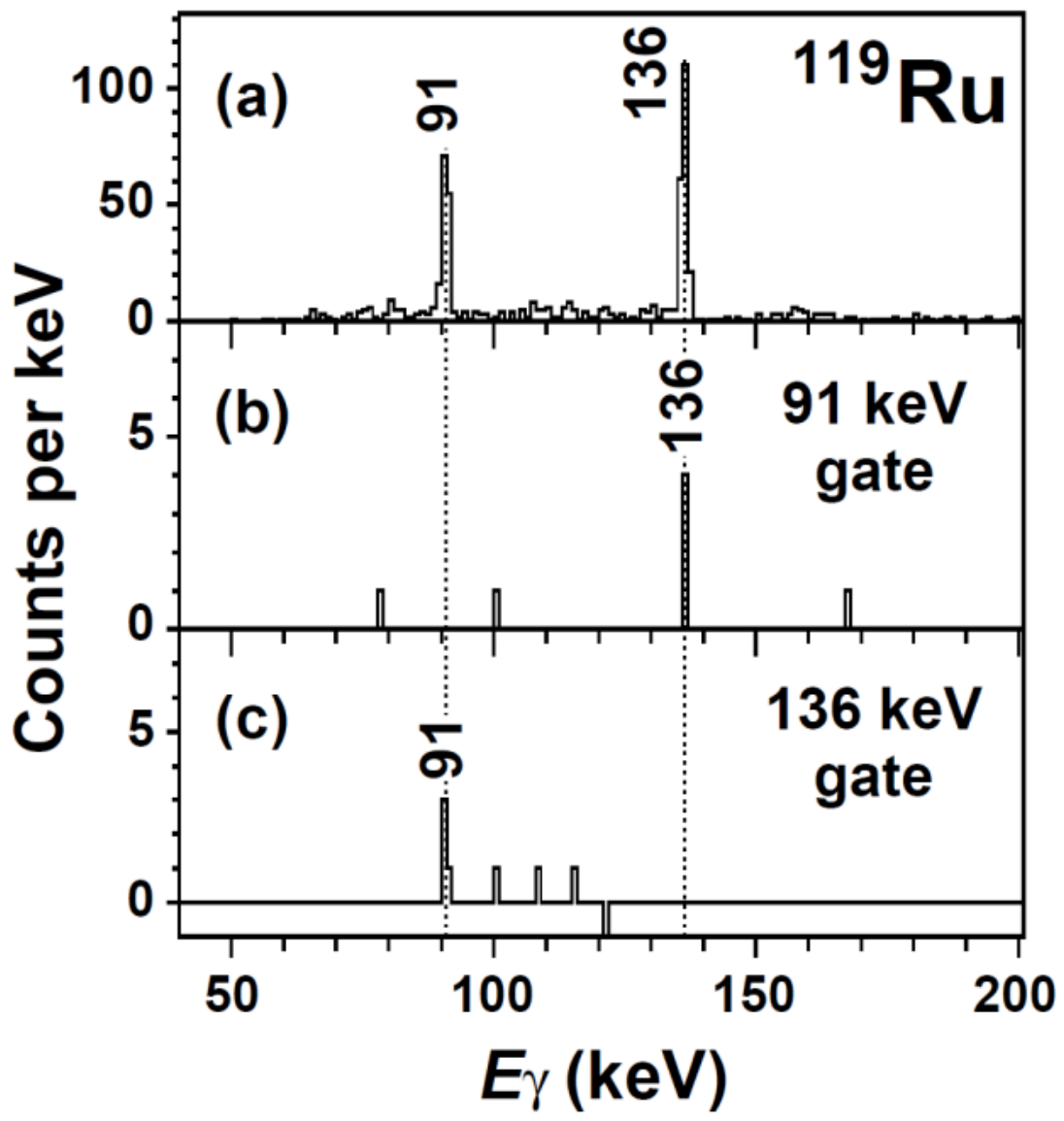

FIG. 5. Delayed $\gamma$-ray singles spectrum and $\gamma\gamma$ coincidence spectra of $^{119}$Ru$^{m}$. The $\gamma$-ray spectra observed in coincidence with (b) the 90.8- and (c) 136.3-keV $\gamma$-rays are shown along with (a) the $\gamma$-ray singles spectrum. Peaks are labeled by energies in keV.

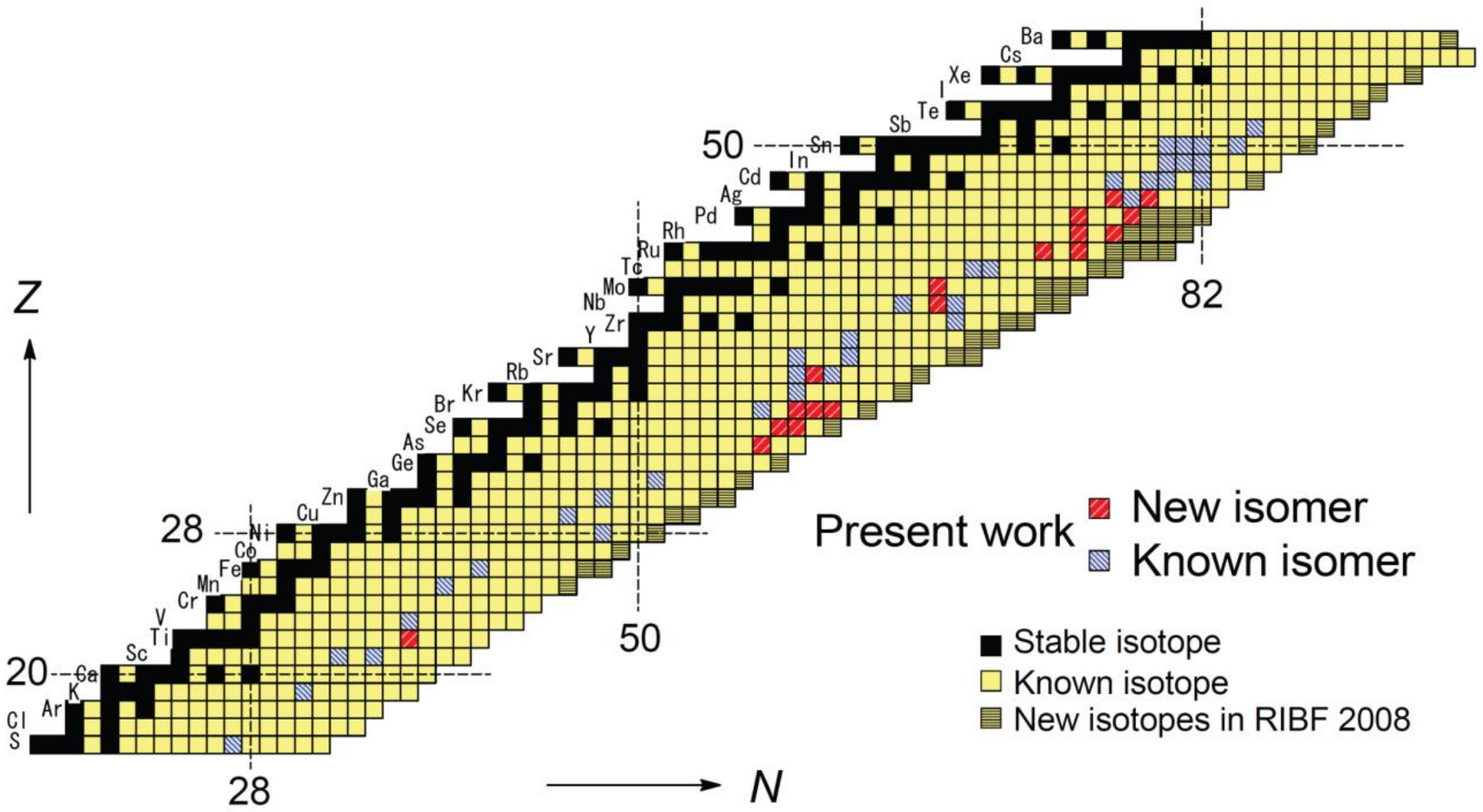

FIG. 6. (Color) Nuclear chart showing the observed isomers in the present work.

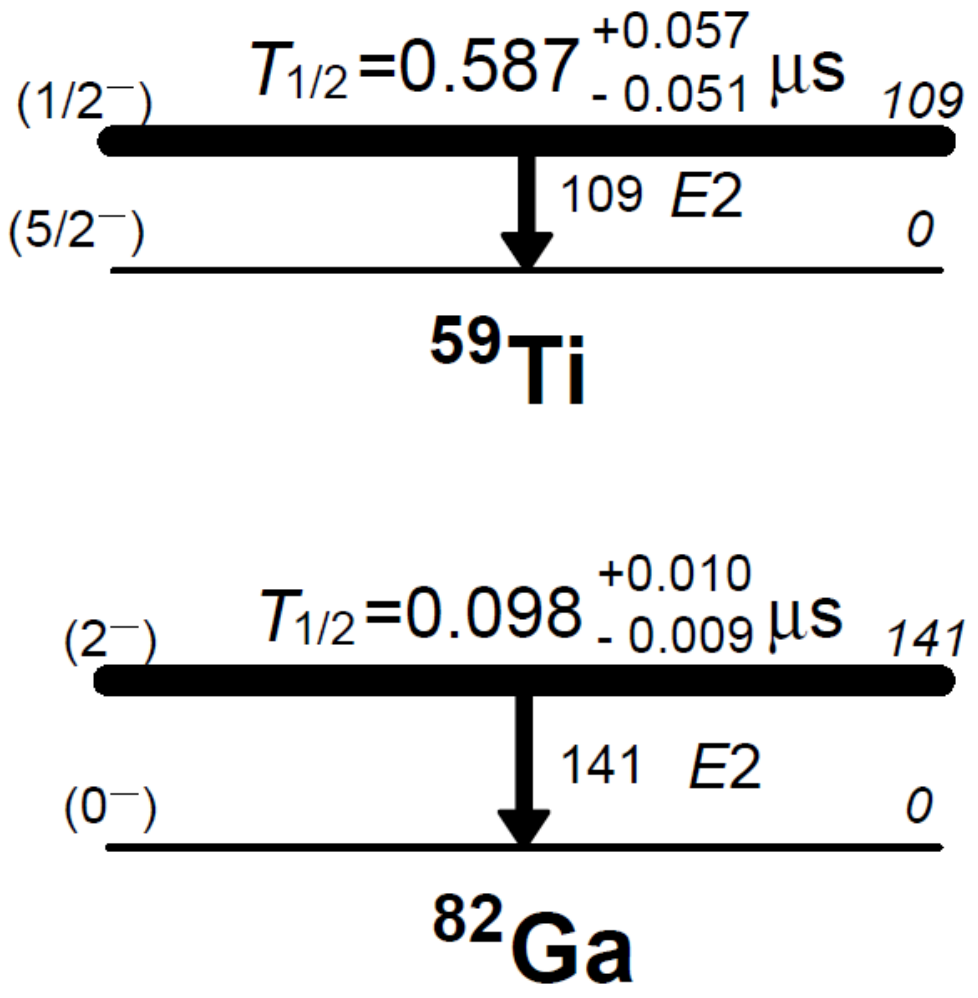

FIG. 7. Proposed level schemes of $^{59}Ti^{m}$ and $^{82}Ga^{m}$.

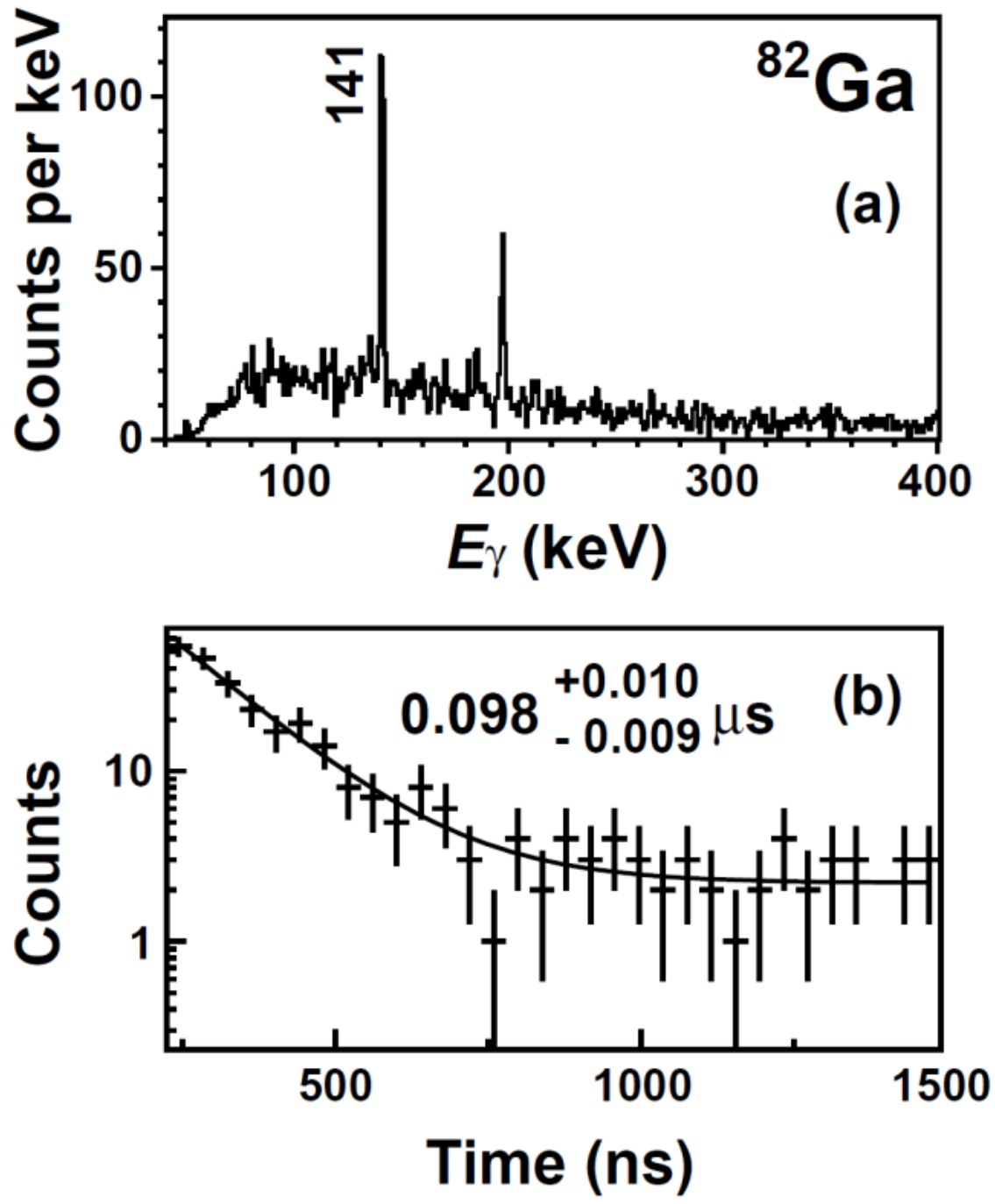

FIG. 8. (a) Delayed $\gamma$-ray energy spectrum and (b) time spectrum of $^{82}Ga^{m}$. The $\gamma$-ray peak at 197 keV is attributed to $^{19}F(n,n'\gamma)^{19}F$. See text. The deduced half-life is given in the figure (b).

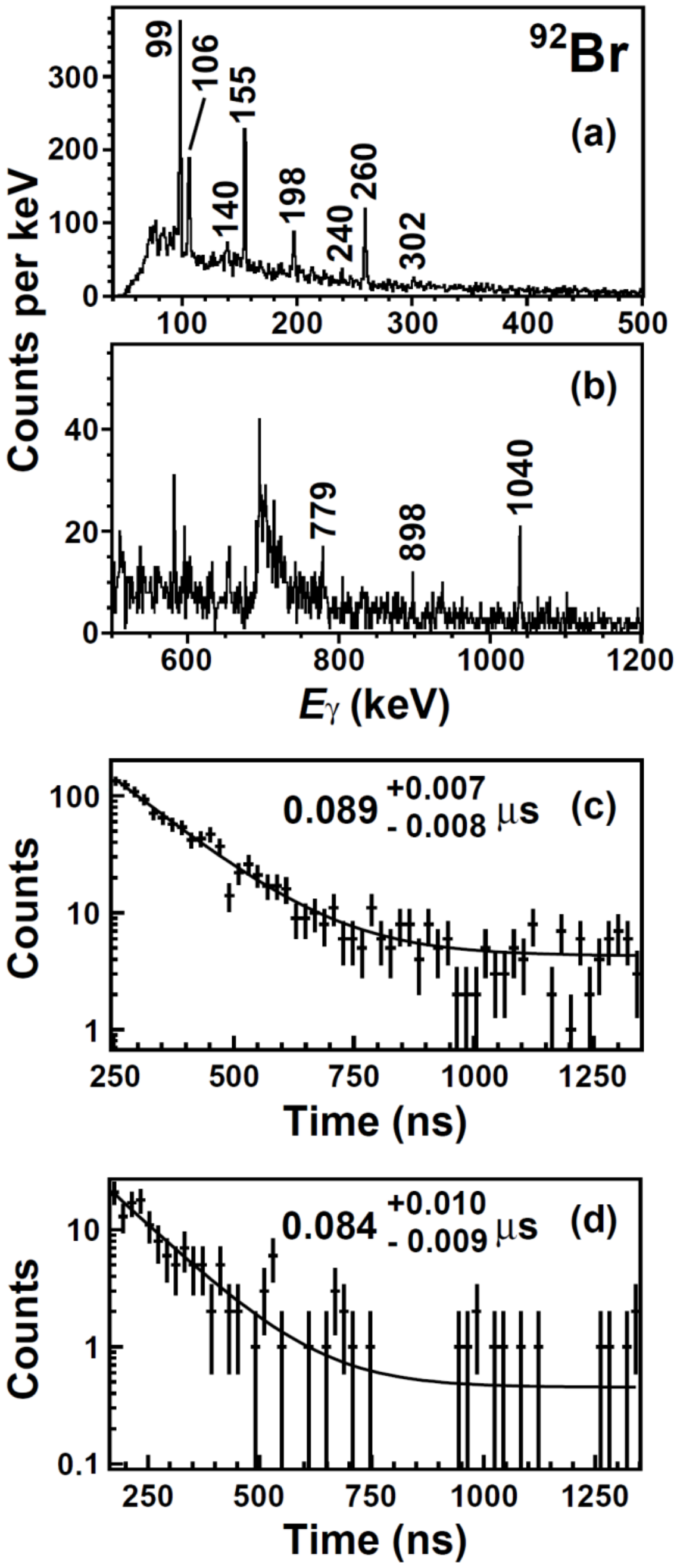

FIG. 9. (a), (b) Delayed $\gamma$-ray energy spectra and (c), (d) time spectra of $^{92}Br^{m}$. The figures (a) and (b) shows the low- and high-energy parts of the energy spectrum, respectively. Peaks are labeled by energies in keV. Unlabeled peaks in the energy spectra are the background $\gamma$-rays described in text or unidentified $\gamma$-rays. The time spectra were obtained using (c) the $\gamma$-rays at 106.3, 154.9 and 301.9 keV for the 662-keV isomer and (d) those at 778.8, 898.3 and 1039.8 keV for the 1138-keV isomer. The deduced half-lives are given in the figures. See text.

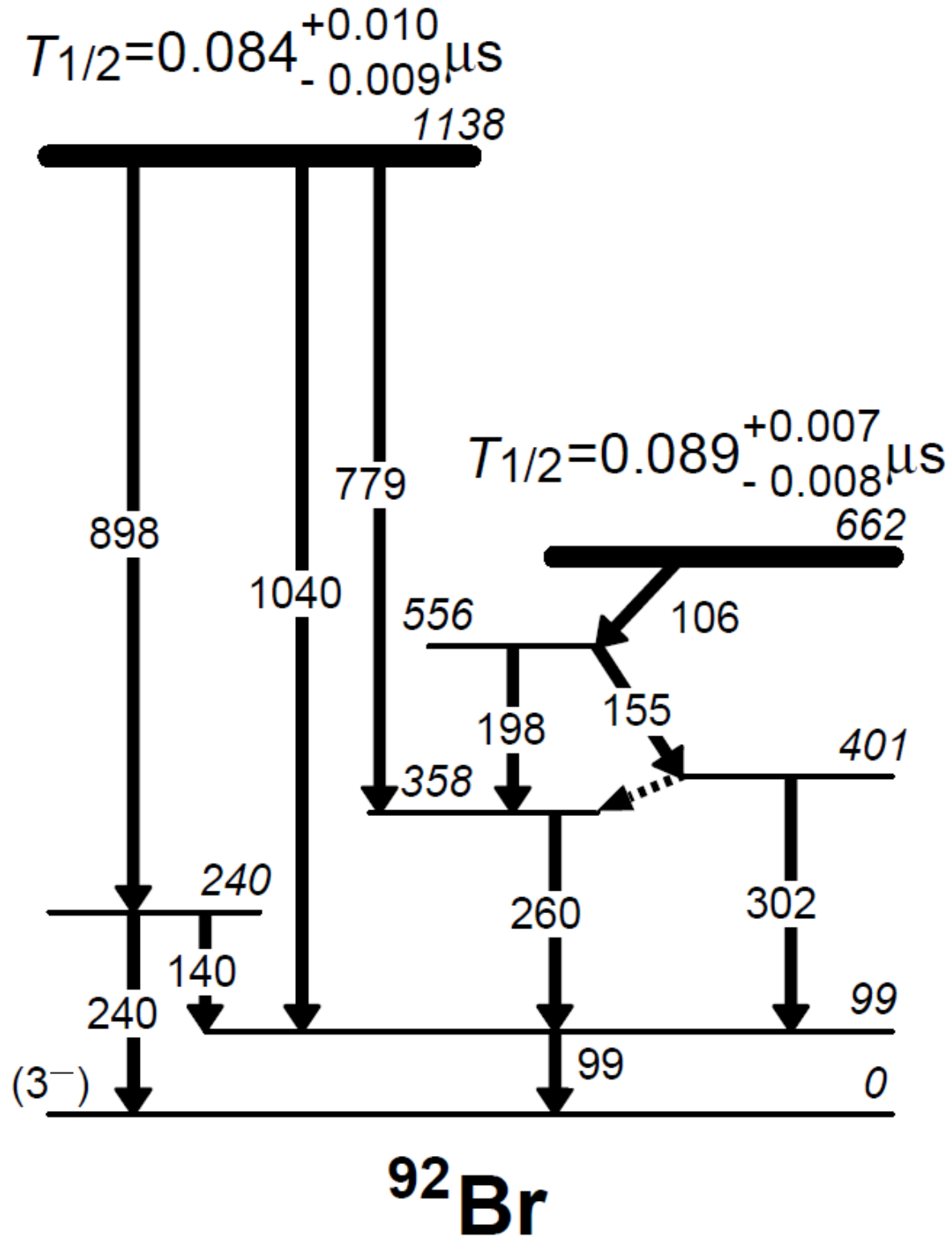

FIG. 10. Proposed level scheme of $^{92}Br^{m}$.

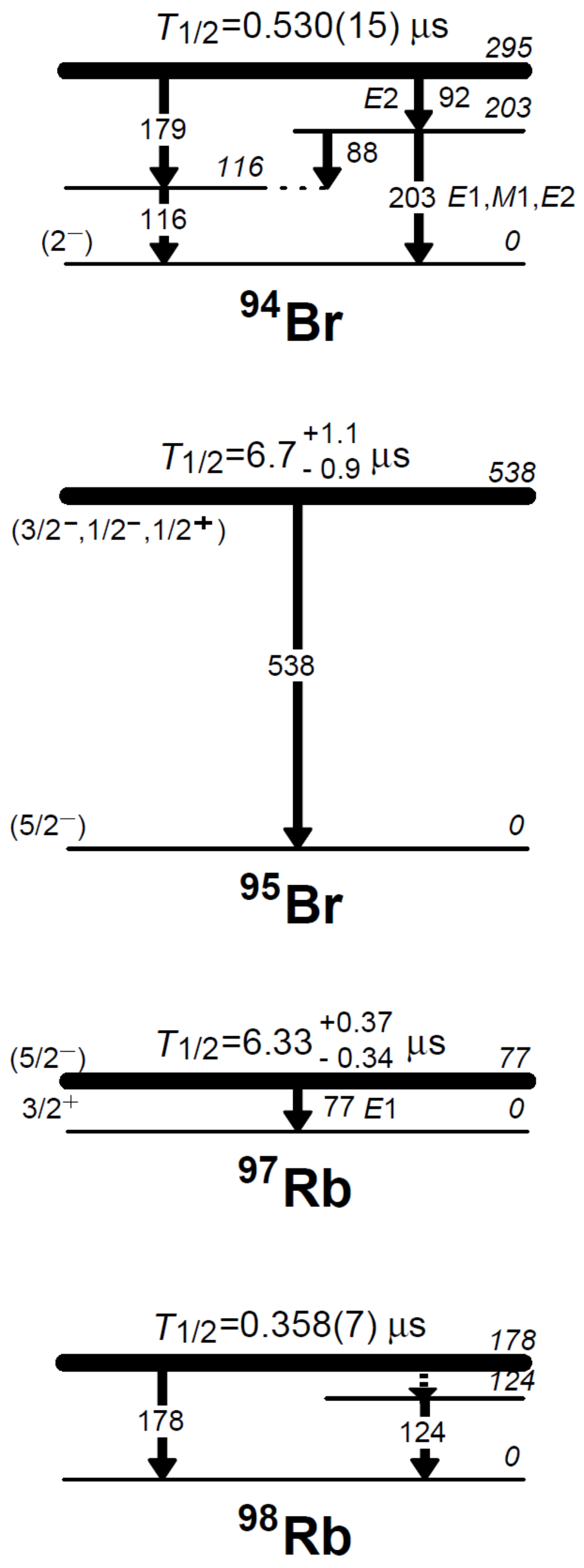

FIG. 11. Proposed level schemes of $^{94}Br^{m}$, $^{95}Br^{m}$, $^{97}Rb^{m}$ and $^{98}Rb^{m}$.

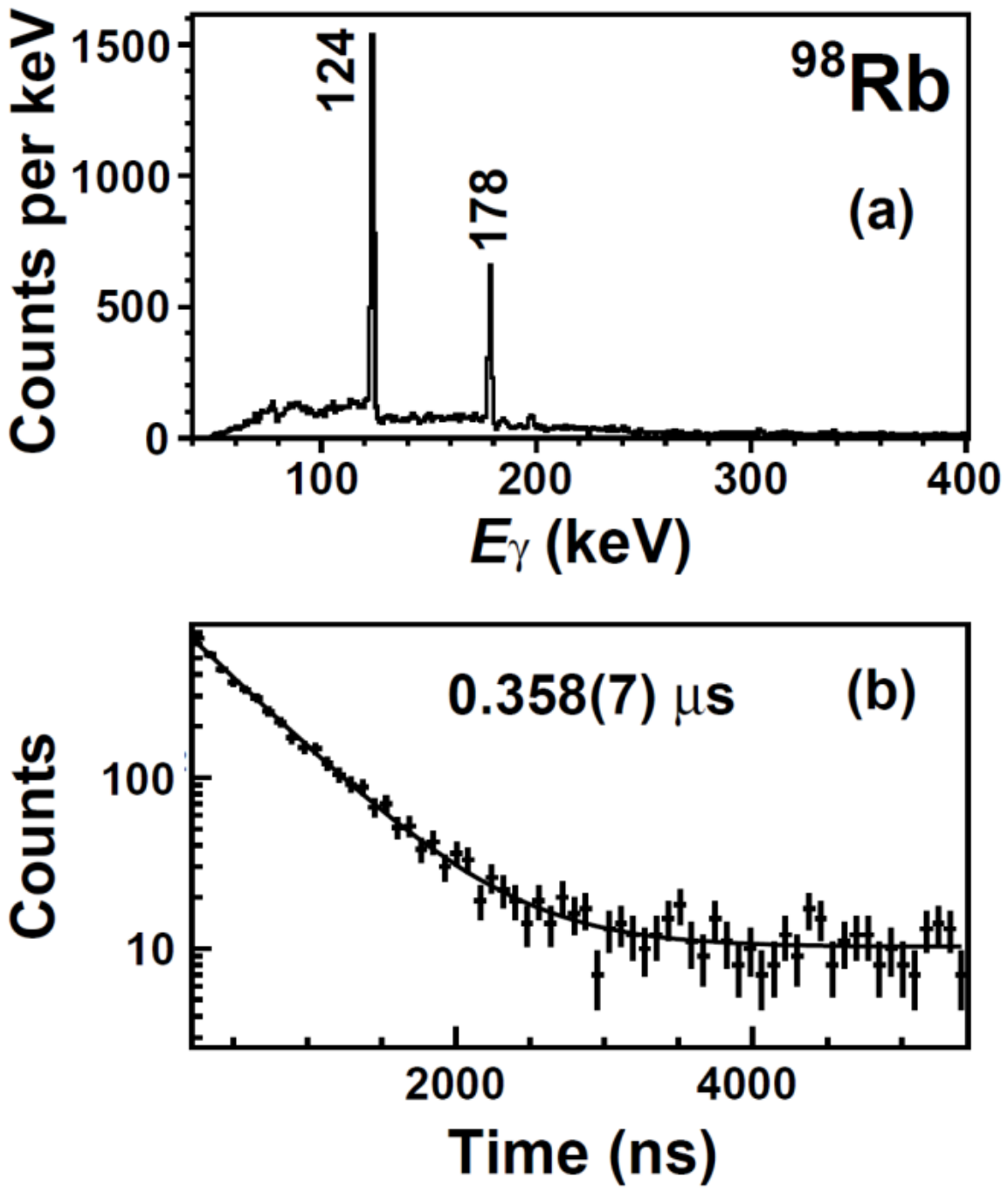

FIG. 12. (a) Delayed $\gamma$-ray energy spectrum and (b) time spectrum of $^{98}Rb^{m}$. Unlabeled peaks in the energy spectrum are the background $\gamma$-rays described in text or unidentified $\gamma$-rays. The deduced half-life is given in the figure (b).

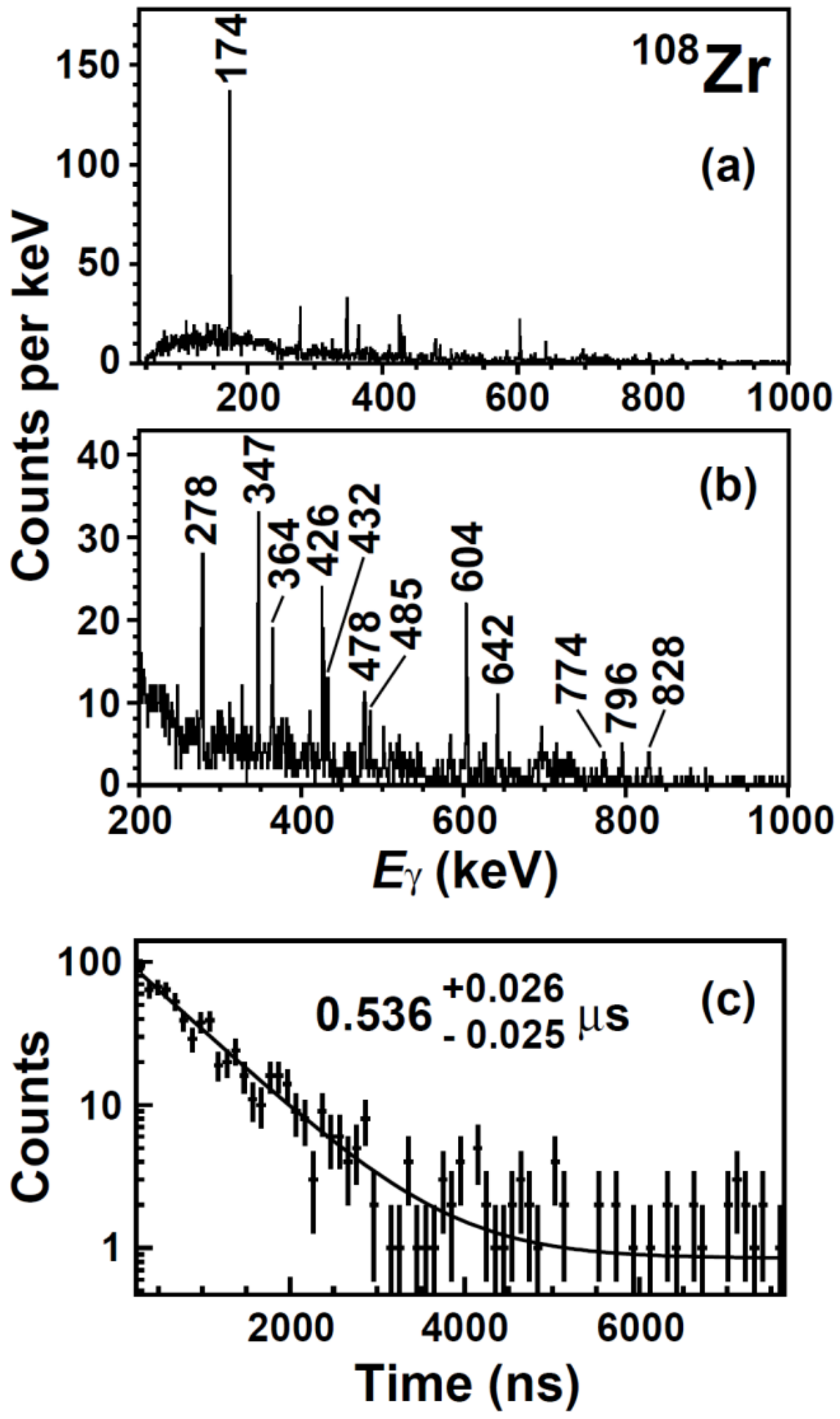

FIG. 13. (a), (b) Delayed $\gamma$-ray energy spectra and (c) time spectrum of $^{108}Zr^{m}$. The figure (b) shows a partially-enlarged energy spectrum. Unlabeled peaks in the energy spectra are the background $\gamma$-rays described in text or unidentified $\gamma$-rays. The deduced half-life is given in the figure (c).

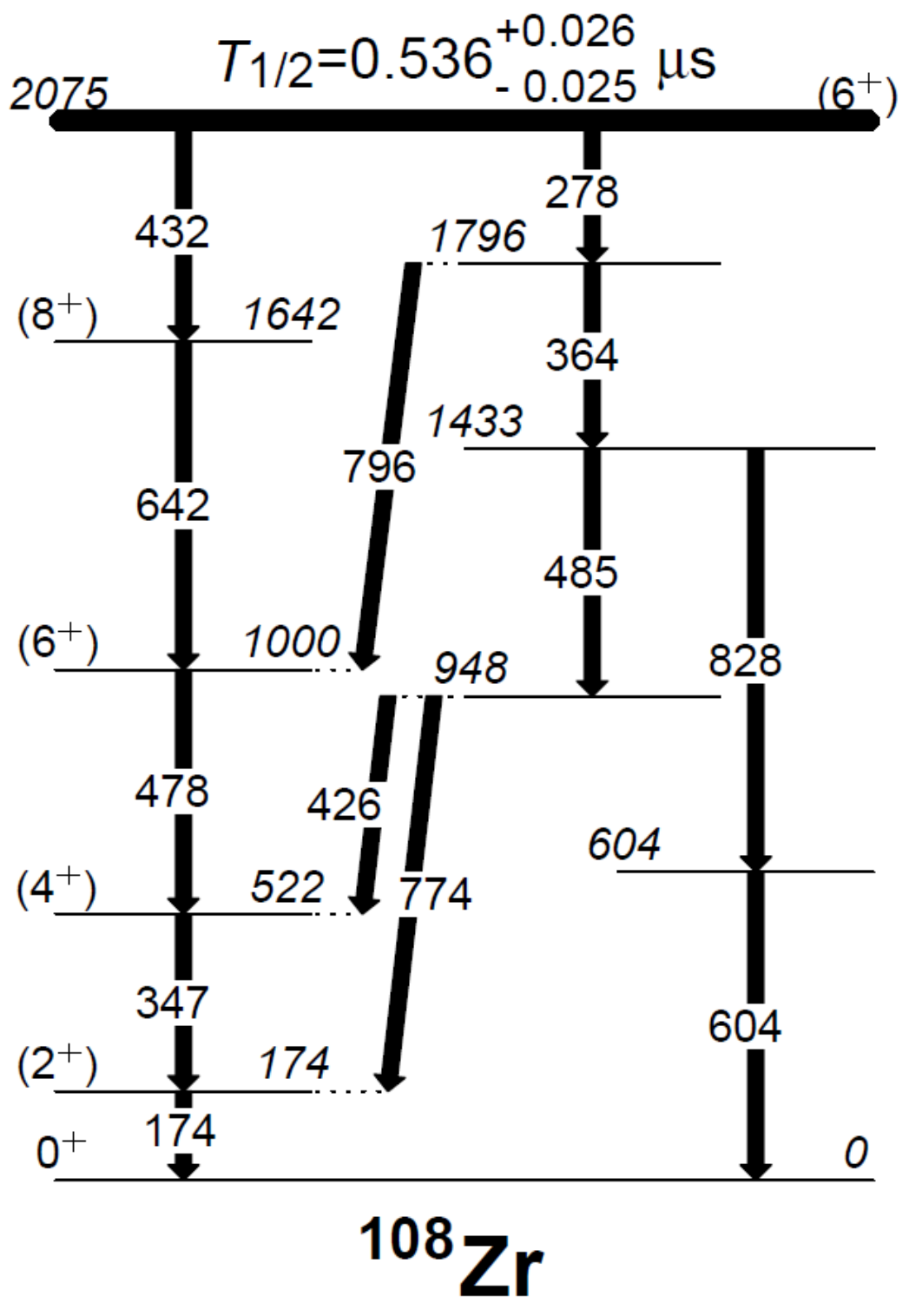

FIG. 14. Proposed level scheme of $^{108}Zr^m$.

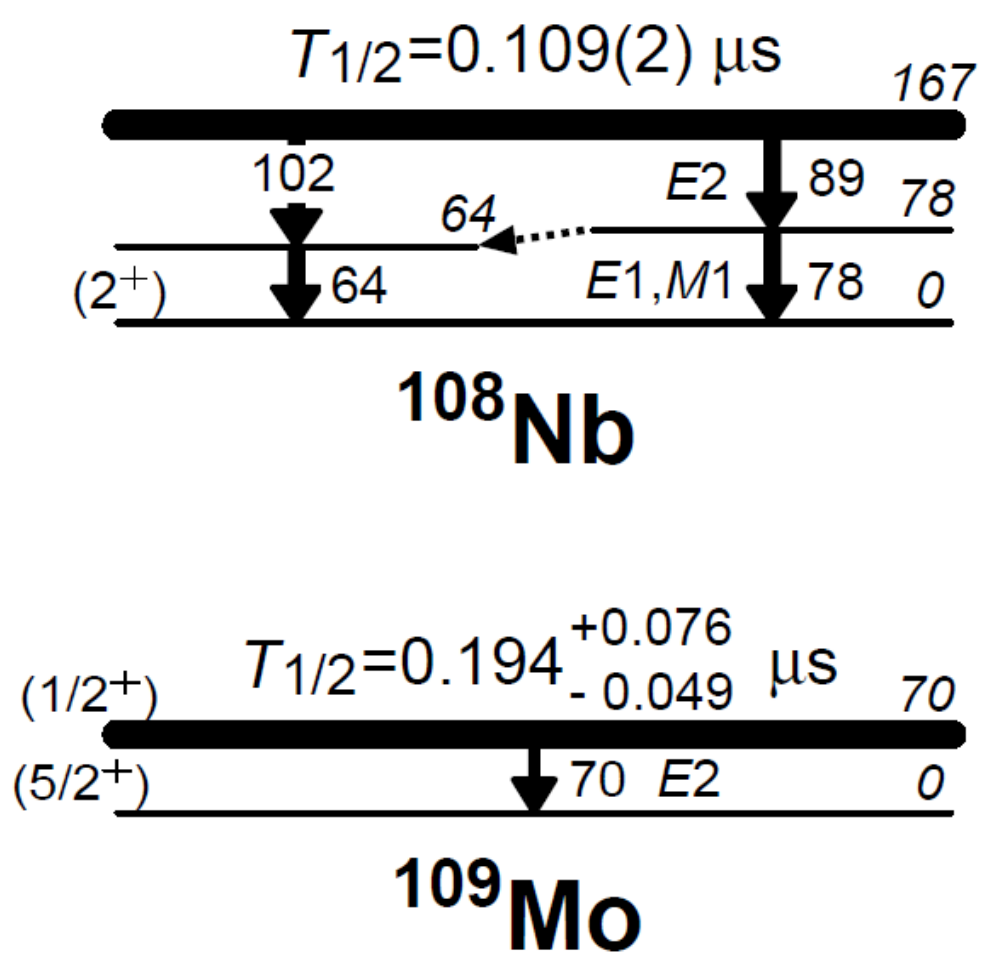

FIG. 15. Proposed level schemes of $^{108}Nb^m$ and $^{109}Mo^m$.

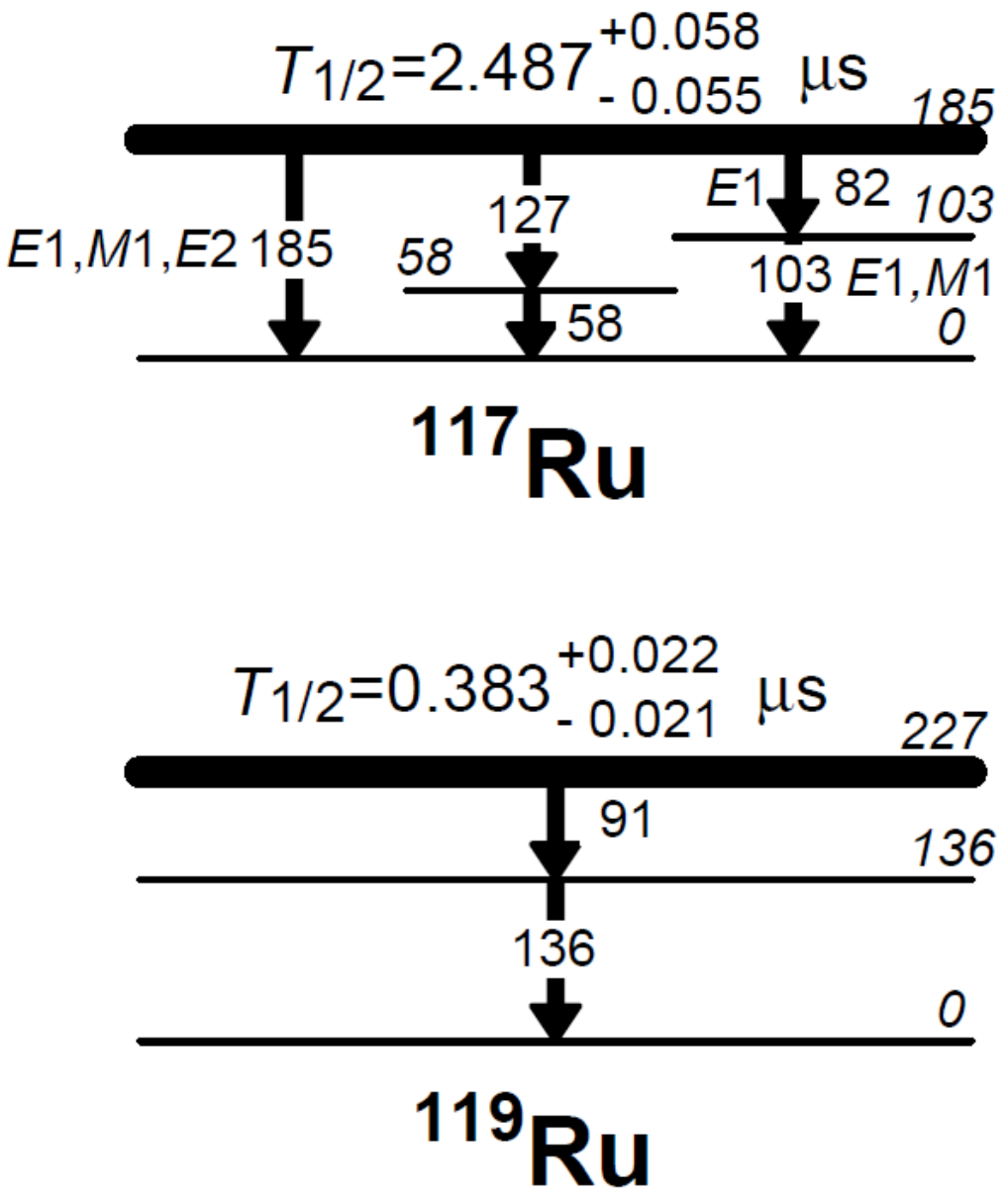

FIG. 16. Proposed level schemes of $^{117}Ru^{m}$ and $^{119}Ru^{m}$.

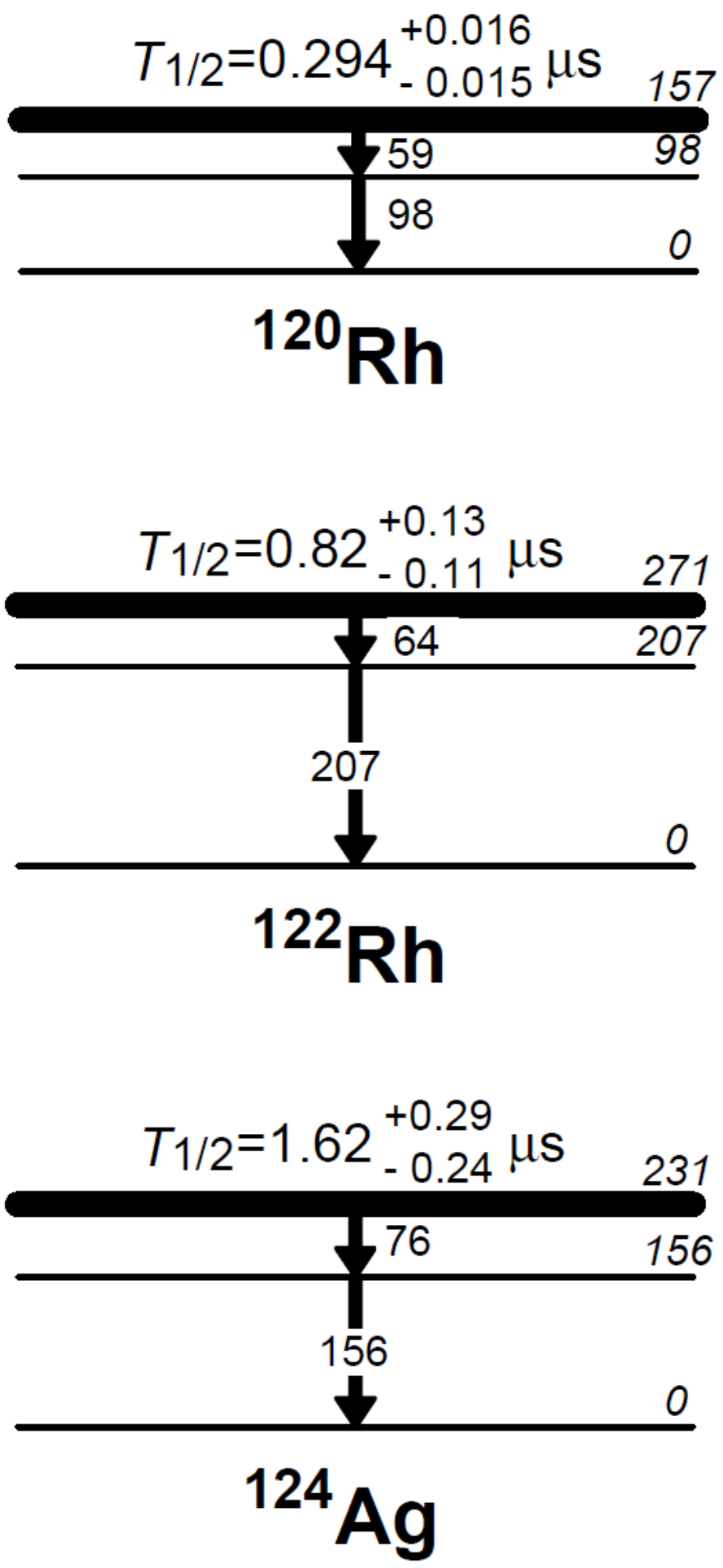

FIG. 17. Proposed level schemes of $^{120}Rh^{m}$, $^{122}Rh^{m}$ and $^{124}Ag^{m}$.

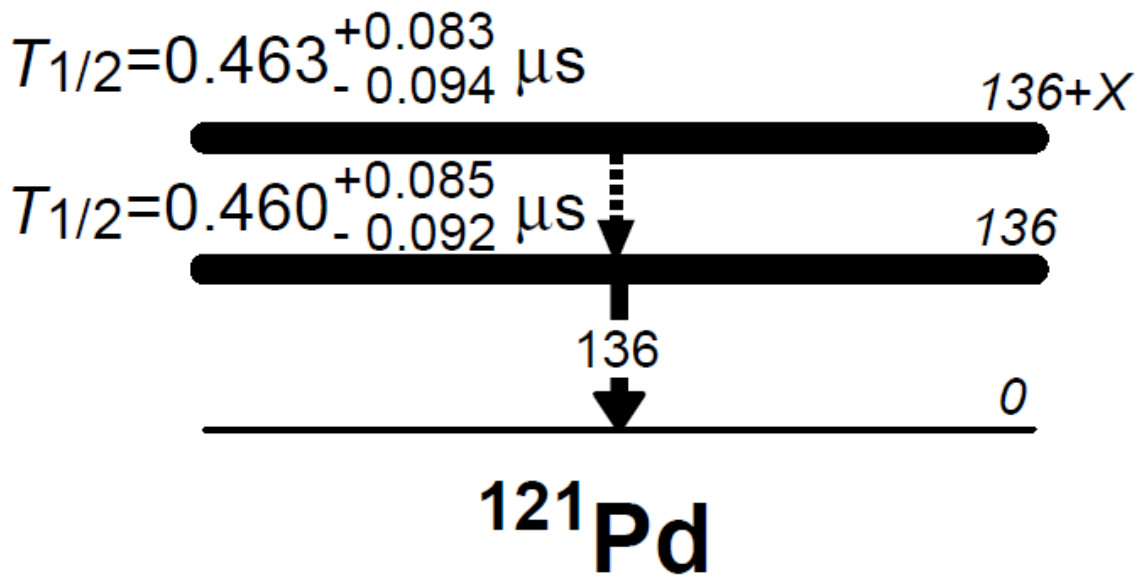

FIG. 18. Proposed level scheme of $^{121}Pd^{m}$.

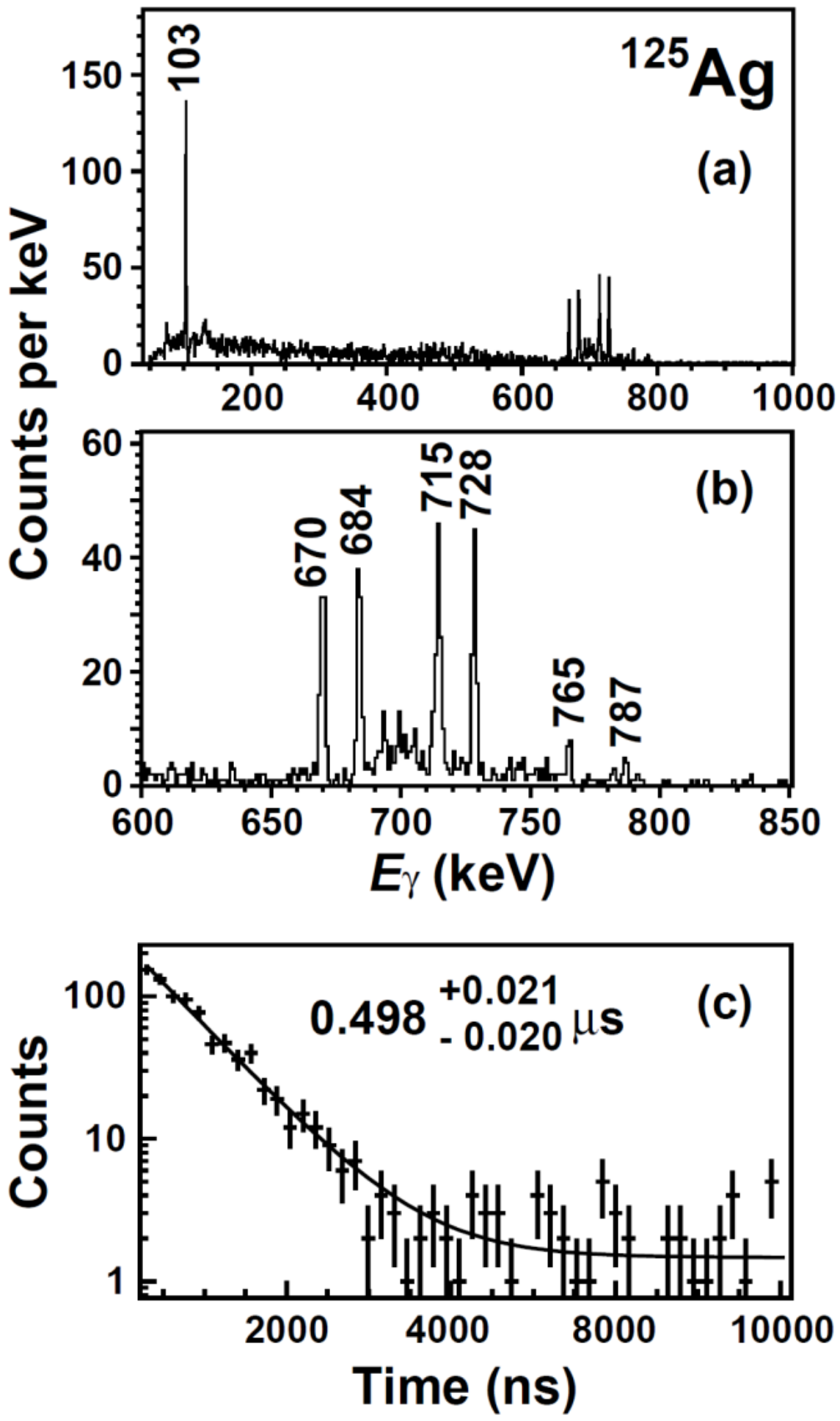

FIG. 19. (a), (b) Delayed $\gamma$-ray energy spectra and (c) time spectrum of $^{125}Ag^{m}$. The figure (b) shows a partially-enlarged energy spectrum. Unlabeled peaks in the energy spectra are the background $\gamma$-rays described in text or unidentified $\gamma$-rays. The deduced half-life is given in the figure (c).

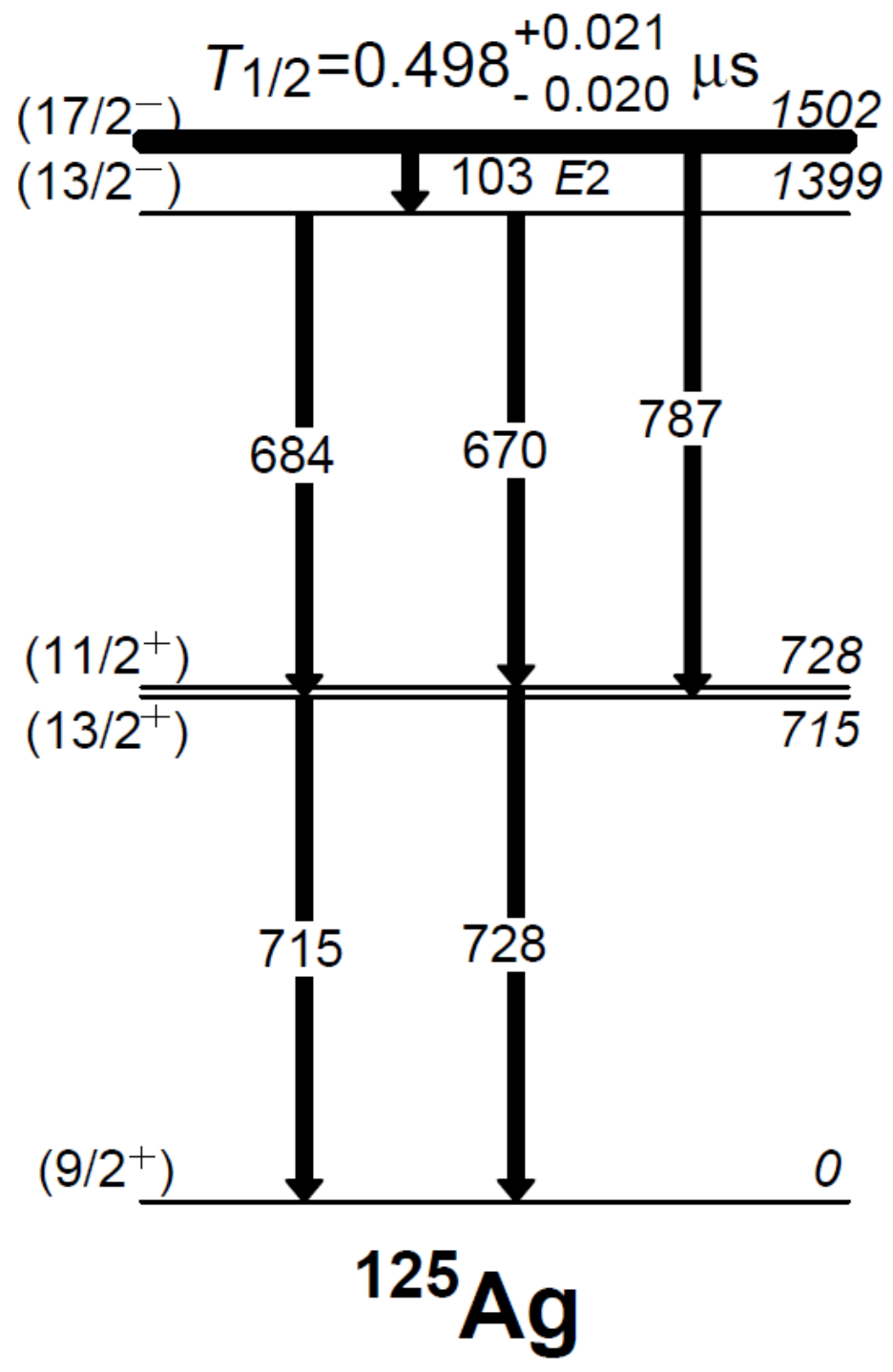

FIG. 20. Proposed level scheme of $^{125}Ag^{m}$.